\documentclass[fleqn,10pt]{wlscirep}
\usepackage[utf8]{inputenc}
\usepackage[T1]{fontenc}
\usepackage{colortbl}
\usepackage{graphicx}
\usepackage{wrapfig}
\usepackage[font=footnotesize,labelfont=bf]{caption}
\usepackage{mathrsfs}
\usepackage[mathcal]{eucal}
\usepackage{multirow}
\usepackage{float}
\usepackage{verbatim}
\usepackage{enumerate}
\usepackage[textsize=tiny, colorinlistoftodos, textwidth=15mm, shadow]{todonotes}

\title{Deep Generative Modeling for Volume Reconstruction in Cryo-Electron Microscopy}

\author[1+]{Claire Donnat}
\author[2,3]{Axel Levy} 
\author[3]{Frédéric Poitevin}
\author[4]{Ellen Zhong}
\author[5*+]{Nina Miolane}
\affil[1]{University of Chicago, Department of Statistics, Chicago, Illinois, USA}
\affil[2]{Stanford University, Department of Electrical Engineering, Stanford, CA, USA}
\affil[3]{LCLS, SLAC National Accelerator Laboratory, Menlo Park, CA, USA}
\affil[4]{Massachusetts Institute of Technology, Computer Science and Artificial Intelligence Lab, Boston, MA, USA }
\affil[5]{University of California Santa Barbara, Department of Electrical \& Computer Engineering, Santa Barbara, CA, USA}
\affil[*]{ninamiolane@ucsb.edu}

\affil[+]{these authors contributed equally to this work}

\DeclareMathOperator*{\argmax}{arg\,max}
\DeclareMathOperator*{\argmin}{arg\,min}

\newcommand{\cryoposenet}{CryoPoseNet\cite{Nashed2021endtoend}}
\newcommand{\cryogan}{CryoGAN\cite{gupta2021cryogan}}
\newcommand{\multicryogan}{Multi-CryoGAN\cite{gupta2020multi}}

\newcommand{\threedflex}{3DFlex\cite{Punjani2021} }
\newcommand{\relion}{RELION\cite{Scheres2012ADetermination} }
\newcommand{\cryomaxwelling}{FSTdiff\cite{Ullrich2019}}
\newcommand{\cryovaegan}{CryoVAEGAN\cite{MiolanePoitevin2020CVPR} }
\newcommand{\cryodrgn}{CryoDRGN\cite{Zhong2019ReconstructingModels} }
\newcommand{\egmm}{E2GMM\cite{Chen2021} }
\newcommand{\cryofold}{CryoFold\cite{zhong2021exploring} }
\newcommand{\cryodeepmind}{AtomVAE\cite{Rosenbaum2021} }
\newcommand{\cryosparc}{CryoSPARC\cite{Punjani2017CryoSPARC:Determination}}
\newcommand{\cryoai}{CryoAI\cite{Levy2022CryoAI}}

\usepackage{graphics}

\usepackage[acronym]{glossaries}

 \makenoidxglossaries

\newglossaryentry{cryoem}
{
    name=cryo-EM,
    description={Cryogenic Electron Microscopy}
}

\newglossaryentry{reconstruction}
{
    name=reconstruction,
    description={In the context of imaging, reconstruction usually refers to tomographic reconstruction of 2D images into a 3D volume}
}

\newglossaryentry{poses}
{
    name=poses,
    description={The unknown 3D orientation and 2D translation of the particle within a given projection image}
}

\newglossaryentry{generativemodel}
{
    name=generative model,
    description={A generative model is a statistical model of the joint probability distribution for the data and the underlying quantity of interest. It is a process that describes how data is generated, using probability distributions. By sampling from this model, we create new data points}
}

\newglossaryentry{deepgenerativemodels}
{
    name=deep generative models,
    description={A class of generative models leveraging deep neural networks}
}

\newglossaryentry{spi}
{
    name=single particle images,
    description={Single Particle Imaging refers to imaging modalities where data, \emph{albeit} corrupted, is collected about individual particles, as opposed to ensemble averaging imaging modalities such as X-ray crystallography or NMR}
}

\newglossaryentry{particles}
{
    name=particles,
    description={In cryo-EM, each copy of an imaged molecule is referred to as a \emph{particle}}
}

\newglossaryentry{conformational-hetero}
{
    name=heterogeneity,
    description={
    Heterogeneity refers to structural variability between the imaged particles, whether from compositional variability in the sample or conformational dynamics associated with the molecule. Heterogeneous reconstruction algorithm aim to model the \gls{conformation landscape} associated with the target molecule}
}

\newglossaryentry{conformation landscape}
{
    name=conformational landscape,
    description={Molecules exhibit shape variability as they thermally diffuse on their high-dimensional energy potential surface, also known as their conformation landscape. Indeed, each atom of the molecule is associated with 6 degrees of freedom, three for their position in space and three for their momentum. The joint distribution over all these degrees of freedom is directly related to the pairwise interaction between atoms which add up to the potential energy of the molecule. The likelihood of a conformation is dictated by the potential energy value associated with this conformation and thus by the relative position of atoms. One could think of the potential energy function as a surface in high-dimension with basins corresponding to stable conformations separated by high energy intermediates. The overall energy scale and relative depths of the basins is dictated by thermodynamics observables such as the temperature; Low temperature will exacerbate differences in energy and differentially increase the likelihood of conformations found in basins relative to higher energy intermediate, for example}
}

\newglossaryentry{colvar}
{
    name=collective variables,
    description={The conformational space of a molecule can be defined by features that are functions of the coordinates of the molecule's atoms, often referred to as collective variables since most interesting features tend to capture global collective motions of the atoms}
}

\newacronym{ctf}{CTF}{Contrast Transfer Function}

\newacronym{psf}{PSF}{Point Spread Function}

\newacronym{ml}{ML}{Maximum Likelihood}

\newacronym{map}{MAP}{Maximum a Posteriori}

\newacronym{em}{EM}{Expectation-Maximization}

\newacronym{elbo}{ELBO}{Evidence Lower Bound}

\newacronym{kl}{KL}{Kullback-Leibler divergence}

\newacronym{vi}{VI}{Variational Inference}

\newacronym{ai}{AI}{Amortized Inference}

\newacronym{gan}{GAN}{Generative Adversarial Network}

\newacronym{fsc}{FSC}{Fourier Shell Correlation}

\newacronym{rmsd}{RMSD}{Root Mean Square Deviation}

\begin{abstract}
  Recent breakthroughs in high-resolution imaging of biomolecules in solution with cryo-electron microscopy (cryo-EM) have unlocked new doors for the reconstruction of molecular volumes, thereby promising further advances in biology, chemistry, and pharmacological research. 
 Recent next-generation volume reconstruction algorithms that combine generative modeling with end-to-end unsupervised deep learning techniques have shown promising preliminary results, but still face considerable technical and theoretical hurdles when applied to experimental cryo-EM images. In light of the proliferation of such methods, we propose here a critical review of recent advances in the field of \textit{deep generative modeling for  cryo-EM volume reconstruction}. The present review aims to (i) unify and compare these new methods using a consistent statistical framework, (ii) present them using a terminology familiar to machine learning researchers and computational biologists  with no specific background in cryo-EM, and (iii) provide the necessary perspective on current advances to highlight their relative strengths and weaknesses, along with outstanding bottlenecks and avenues for improvements in the field.
 This review might also raise the interest of computer vision practitioners, as it highlights significant limits of deep generative models in low signal-to-noise regimes --- therefore emphasizing a need for new theoretical and methodological developments.
\end{abstract}

\begin{document}

\flushbottom
\maketitle

\thispagestyle{empty}

\setlength{\belowdisplayskip}{1pt} \setlength{\belowdisplayshortskip}{1pt}
\setlength{\abovedisplayskip}{1pt} \setlength{\abovedisplayshortskip}{1pt}

\section*{Introduction}\label{sec0-intro}

\begin{wrapfigure}{r}{0.4\textwidth}
\vspace*{-0.2in}
    \includegraphics[width=0.4\textwidth]{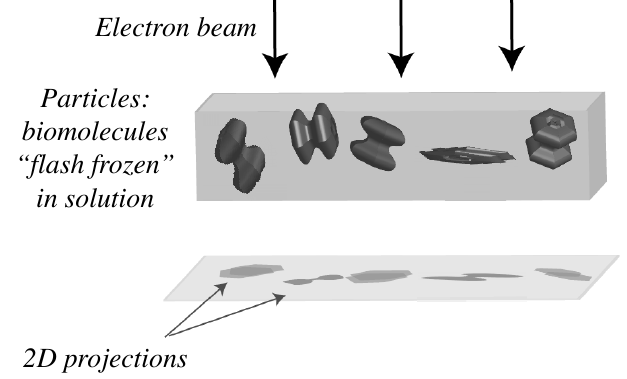}
     \setlength{\belowcaptionskip}{-4pt} 
      \setlength{\abovecaptionskip}{-4pt} 
    \caption{Acquisition of 2D cryo-EM images (2D projections) from 3D biomolecular volumes.}
    \label{fig:electron_microscope}  \vspace*{-0.08in}
\end{wrapfigure}

High-resolution reconstruction of molecular volumes from \gls{spi} has the potential to facilitate new breakthroughs in our ability to understand fundamental biological mechanisms and engineer macromolecular function \cite{Ourmazd2019,Renaud2018cryo}. In this context, cryo-electron microscopy (\gls{cryoem}) has fostered a revolution in structural biology by allowing the imaging of biomolecules in solution at atomic resolution \cite{Nakane2020atomicres,emdb}.
However, the estimation of these molecules' 3-dimensional (3D) volume from cryo-EM data continues to pose a formidable challenge. In this setting, observations are limited to the raw 2D projections of molecules  (also called \gls{particles}) relative to an incoming electron beam, while their 3D orientation and position (jointly called \gls{poses}) are unknown --- see Figure~\ref{fig:electron_microscope}. Reconstructing molecular volumes therefore also requires recovering a number of hidden variables such as each particle's 3D orientation. The difficulty of this task is further compounded by a combination of factors, including the variability in the shape of any given molecule (also referred to as structural ``\gls{conformational-hetero}''), the non-linear physics of the data acquisition process, as well as extremely low signal-to-noise ratios --- concepts formalized in the image formation model below.

 \paragraph{Image Formation Model.}
The process of image formation in cryo-EM involves several physical phenomena, including pairwise interactions between atoms, interactions between the electron beam and the molecule's electrostatic potential, or microscope effects. 
We refer the reader to Dill \textit{et al.}\cite{dill2010molecular}, Kohl and Reimer\cite{Kohl}, and Vulovic \textit{et al.}\cite{VULOVIC201319} for in-depth descriptions of these phenomena. Nonetheless, in most cases\cite{Scheres2012RELION,VULOVIC201319}, each image $X_i$ in a dataset of  $n$ images of single \gls{particles}
can be modeled as a random sample from the following \gls{generativemodel}:
\begin{equation}\label{eq:image}
    X_i = \text{PSF}_i * (t_i \circ \Pi_{2D} \circ R_i) (V^{(i)})  + \epsilon_i, \qquad \text{with} \quad i=1 \cdots n.
\end{equation}
\noindent Here, $R_i$ is a 3D rotation representing the 3D orientation of the volume $V^{(i)}$ with respect to the direction of the electron beam. The oriented volume is subsequently ``pierced through'' by the electron beam and projected onto the detector --- an operation represented in Equation~\eqref{eq:image} by the 2D-projection operator $\Pi_{2D}$. The  variable $t_i$ represents the 2D translation of the projected volume with respect to the center of the image. 
The effect of the microscope's lens is modeled through the convolution $*$ of the 2D projection by an image-dependent operator $\text{PSF}_i$ called the Point Spread Function (\acrshort{psf}) of the microscope whose parameters can depend on the image. Finally, additional noise $\epsilon_i$ is introduced in the observed image, and typically assumed to be Gaussian with zero mean and variance $\sigma_i^2$. Note that the underlying volume $V^{(i)}$ is allowed to depend on $i$. This allows us to account for ``conformational heterogeneity'', a concept whereby a molecule does not necessarily exist in a single state, but rather, that its volume corresponds in fact to one of several stable geometries that can be achieved by this  molecule.

An equivalent generative model can be formulated via the Fourier transform of Equation~\eqref{eq:image} which we present in Appendix~\ref{appB-operators}. In Fourier Space, the convolution $*$ of the projected volume by the PSF becomes a computationally lightweight element-wise matrix multiplication $\odot$ between the 2D Fourier transform of the projected image and that of the PSF (known  as the {\it Contrast Transfer Function}, or {\it CTF}). Operating in Fourier space is thus common in many cryo-EM volume reconstruction algorithms.


\paragraph{Challenges of Cryo-EM and Deep Generative Modeling.} Whether in image or Fourier space, the image formation model described by Equation~\eqref{eq:image} gives us the first clue about the difficulty of the reconstruction problem. \textit{Only the recovery of the set of possible 3D molecular volumes $V^{(i)}$  (called the  \gls{conformation landscape}) is relevant - yet the image also depends on a number of additional unknown nuisance variables}, such as
the pose  ($R_i$, $t_i$) or the microscope-dependent $\text{PSF}_i$. As noted by Singer \textit{et al.} \cite{singer2020computational}, these nuisance variables do not bring any information of biological relevance but may impact the quality (and difficulty) of the reconstruction. Reconstructing molecular volumes thus becomes a highly non-convex optimization problem, putting algorithms at risk of being overly sensitive to initialization \cite{Rosenbaum2021} or converging to one of many local minima \cite{boyd2004convex}. The difficulty of the task is also compounded by the high levels of radiation damage, structured (non-white) noise and altogether remarkably low signal-to-noise ratios that cryo-EM images usually exhibit. Given these difficulties, reconstruction algorithms rely on using copious amounts of data, but often struggle to process the wealth of incoming images\cite{Kimanius2016AcceleratedRELION-2}, to the point where ``the cost of image analysis can exceed 500,000 CPU hours on large, expensive computer clusters''\cite{Punjani2017CryoSPARC:Determination} for a single experiment. In this context, recent efforts have turned to unsupervised deep learning for cryo-EM reconstruction, \textit{i.e.} approaches allowing significant speed-ups through the use of gradient methods and GPUs. Given their potential to advance the field by addressing the challenges mentioned here, we propose here a critical review of deep generative modeling for cryo-EM volume reconstruction.

\paragraph{Related Work.} Several reviews have already begun surveying  challenges and advances in cryo-EM reconstruction.
The reviews by Singer \textit{et al.}\cite{singer2020computational} and Bendory \textit{et al.}\cite{bendory2020single} provide a complete description of cryo-EM reconstruction, 
but focus on mathematical foundations of general computational methods, rather than specifically on deep learning approaches. 
Reviews by Si \textit{et al.}\cite{si2021artificial}, Ede \textit{et al.} \cite{ede2021deep} and Wu \textit{et al.} \cite{WuReview2021} describe the pervasiveness of deep learning methods along all steps of the cryo-EM pipeline, without specialising  to high-resolution volume reconstruction. By contrast, our review is a deep dive into the most recent deep generative models for cryo-EM reconstruction.

\paragraph{Objective and Contributions.} Our objective is to shed light into the similarities and differences among recent state-of-the-art, deep-generative reconstruction methods, which we classify according to {\it(i) \it their parametrization of the generative model} (Section~\ref{sec1-generative_models_and_lvms}) and {\it(ii) the inference tools} deployed to fit this generative model (Section~\ref{sec2-inference}). This unification of recent works along a consistent statistical framework allows us to highlight trends, outstanding challenges, and avenues for improvements in the field (Section \ref{sec3-discussion}). We draw insights from our experiences with these approaches to provide an account of the hurdles and potential difficulties that arise in their deployment to real cryo-EM data. Finally, through this review, we hope to catalyze deep learning advances by providing machine learning practitioners and computer vision experts a thorough overview of the challenges that are unique to cryo-EM. 

\section{Generative Modeling for Cryo-EM}\label{sec1-generative_models_and_lvms}

The objective of cryo-EM imaging algorithms is to produce a 3D reconstruction of a given molecule from a dataset of images $\{X_i\}_{i =1 \cdots n}$, where each image corresponds to a ``2D projection'' of the molecule at a different (unknown) orientation and position (Figure~\ref{fig:electron_microscope}). A fundamental hurdle to this objective lies in the fact that each molecule has its own unknown conformation (or shape). Methods that account for this \gls{conformational-hetero} are called \textbf{heterogeneous reconstruction methods}, and can typically aspire to higher resolution reconstructions --- often at the cost of more involved and expensive computations. Conversely, \textbf{homogeneous reconstruction methods} neglect this shape variability and usually represent the \gls{conformation landscape} as a unique molecular volume. This section reviews how choices both in accounting for conformation heterogeneity and in parametrizing the volume yield different formulations of the cryo-EM image formation model.

 \subsection{Conformation Variable $z$.}\label{confvarz}
 
Heterogeneous reconstruction methods introduce an additional variable $z_i$ for each image $i$ within the formation model of Equations~\eqref{eq:image}, which we call the conformation variable. Depending on whether conformation heterogeneity is modeled through a discrete number of states or as a continuous variable, the conformational landscape can be encoded as a discrete family of volumes $\mathcal{V}=\{V_z,~z \in\{1,\ldots,K\}\}$ \textbf{(discrete heterogeneity)}, or as a continuous family $\mathcal{V}=\{V(z),~z \in\mathbb{R}^L\}$ for some integer $L$ \textbf{(continuous heterogeneity)}\cite{Jonic2017}. In both cases, the family is indexed by the variable $z$. We use the notations $V_z$ and $V(z)$ interchangeably. Homogeneous reconstruction can in fact be taken as the special case where $\mathcal{V}$ only comprises a single volume, so that $K=1$ or $L=0$ (i.e. forcing $z=0$). We now write $V^{(i)} = V(z_i)$ in Equation~\eqref{eq:image}.

\paragraph{Interpretation of the Conformation Variable}
From a statistical mechanics perspective, the conformation variable $z_i$ encodes the location of any given single particle along the conformational landscape\cite{dill2010molecular}. For example, if $z_i\in \mathbb{R}$, $z_i$ can be used to sort conformations along a ``conformation path'', that is, a sequence of small transformations that would interpolate two main preferred, dynamically stable states (or conformations) for the molecule. When continuous, the dimension $L$ of this cursor variable $z$ could in principle take any value between 0 (no heterogeneity) and $O(N)$, with $N$ the number of atoms in the molecule. However, two factors tend to drastically limit the number of dimensions of $z$. First, most of the main global dynamics of a molecule are captured by a few \gls{colvar} associated with its low-frequency motion, effectively averaging out a lot of the effects of the high number of degrees of freedom associated with faster motions \cite{Noe2017}. Second, imaging conditions often reduce the ability to resolve the remaining motions, thus reducing the effective dimensionality of $z$\cite{Katsevich2015}. 
In other words, limits in the imaging technology itself restrict the dimension of the variable $z$. In the case of discrete heterogeneity, $z \in \{1, .., K\}$ is an index of minimum energy wells (conformations) in the conformational landscape. Imaging conditions also reduce the ability to resolve too many metastable states, thereby restricting practitioners to choose a low value for $K$.

\paragraph{Discrete vs Continuous Conformational Heterogeneity: Pros, Cons and Discussion.}
Discrete heterogeneity has a rich history in cryo-EM. Popularized by the Class3D\cite{Scheres2012RELION, SCHERES2010295} extension of \relion, it offers the advantage of delivering readily interpretable results: a set of $K$ volumes, representing $K$ main stable states of the molecules. Discrete heterogeneity is thus particularly adequate in certain (common) scenarios where the conformation landscape has local energy minima that produce distinct states. However, one of the main drawbacks of this method consists in the necessary selection of the number $K$ of appropriate conformations. Theoretically, this could be done by cross-validation. In practice, due to the significant computing costs that cross-validation implies, $K$ is chosen in an ad hoc fashion by the experimenter and rarely motivated by strong quantitative arguments (see Haselbach \emph{et al}\cite{Haselbach2018} for a rare example). 

Consequently, many recent methods have turned to a continuous representation of heterogeneity which does not require specifying a number $K$ of conformations. This representation is also often deemed to be closer to the underlying biology, as molecules do not exist as finite/discrete sets of shapes. Rather, a more realistic analogy is to think of molecules as random samples from the equilibrium distribution over their conformational space\cite{dill2010molecular}. However, while a continuous representation could be more scientifically relevant, it remains to be determined how accurate the reconstruction of the conformational space by the space indexed by $z$ truly is. This latter point will be critical to address for heterogeneous reconstruction methods to become more quantitative and directly comparable to other measures from biophysicists and biochemists. We discuss in section \ref{sec3-discussion} the challenges of assessing the precision of such approaches, which probably constitutes one of the main open questions in the field. Additionally, despite its initial appeal, continuous conformation heterogeneity comes with significant theoretical and practical caveats. From a physics perspective, it is still unclear whether the full landscape (at room temperature) is sufficiently well sampled by cryo-EM to justify modeling conformations with a continuous rather than discrete distribution: the sample preparation process in cryo-EM --- and most specifically the grid-freezing step--- affects the distribution of conformations which might not reflect the heterogeneity of conformations at room temperature\cite{Bock2022}. From a statistical perspective, using a continuous distribution necessitates the generative model to be able to sample from the full conformation landscape, a requirement that is itself a considerable challenge for large molecules: the strong constraints, e.g. on on bond lengths and torsion angles, make up for a complex, non-convex landscape that is difficult to sample from. 
Despite these caveats, Table~\ref{tab:volumes} shows that continuous heterogeneity is gaining traction amongst the most recent reconstruction advances.

\subsection{Molecular Volume $V(z)$}\label{definingV} 
The cryo-EM reconstruction problem can thus be understood as recovery of the underlying \gls{conformation landscape} $\mathcal{V}$ and the corresponding probability distribution. The next critical step thus consists in finding a judicious parametrization for each volume $V(z) \in \mathcal{V}$. This requires choosing first an ``output space'' (image space vs. Fourier space, inducing real vs complex values), second, an ``encoding style'' (reference-free vs. reference-based) and third, and ``input domain'' (continuous vs. discrete).

\subsubsection{Defining the output space: image space or Fourier space}
Equations~\eqref{eq:image}-\eqref{eq:fourier} show that the image formation model can be described equivalently in image space or Fourier space. Thus, each volume within the family of conformations $\mathcal{V}$ can be described either in terms of its pixel intensities or its Fourier coefficients. In either case, the volume $V(z)$ associated to the conformation variable $z$ is defined on an input domain $\Omega\subset\mathbb{R}^3$ (the space of coordinates) and outputs values in an output space that is either $\mathbb{R}$ for pixel intensities representing the electron scattering potential of the molecule, or $\mathbb{C}$ to encode the amplitude and phase of the Fourier coefficients. 

\paragraph{Image versus Fourier space: Pros, Cons and Discussion.}{
From a practical standpoint, the choice of the output space is guided by the set of properties and constraints that the analyst wishes to use to guide volume reconstruction. Historically, the Fourier approach has been preferred.
As summarized by Punjani \textit{et al.} \cite{Punjani2021}, working in Fourier space has the benefits of (a) reducing the computational cost of the image formation model (see discussion of Equation~\ref{eq:fourier}), and (b) allowing closed-form maximum likelihood reconstructions when molecules' orientations and positions are known. However, recent methods such as \threedflex have favored image space, where constraints (e.g. smoothness of the deformation, conservation of energy, etc.) are more interpretable and where operations such as interpolation and deformation of the molecule's density map are more naturally parametrized --- whereas the same operations require a careful treatment in Fourier space. For example, interpolation in Fourier space can introduce unwanted artifacts. As highlighted in Table \ref{tab:volumes}, image space computations constitute a promising and increasingly popular avenue for future developments in cryo-EM reconstruction.
}

\subsubsection{Defining an encoding: reference-free or reference-based volume}

The next step lies in the choice of an ``encoding" for the volume $V(z)$. Cryo-EM analysts typically have two choices: {\it (i) using a reference-based parametrization}, which encodes the conformation landscape through its deviation $\Delta V(z)$ from a reference conformation $V_0$, such that $V(z) = V_0 + \Delta V(z)$; or {\it (ii) using a reference-free parametrization} which directly describes each $V(z)$, for instance as a set of atomic coordinates or a low-dimensional embedding, but with no notion of ``reference'' conformation.  

\paragraph{Reference-based versus reference-free: Pros, Cons and Discussion}
If the  column ``Reference Volume" of Table~\ref{tab:volumes} reflects the historical popularity of reference-free encodings, the most recent methods relying on deep-learning seem to have favored a reference-based approach. For instance, \egmm first learns a reference $V_0$ called the ``neutral representation" which then serves in a reference-based encoding of $V(z)$ to further refine the reconstruction by accounting for conformational variability. In \cryodeepmind, Rosenbaum \textit{et al.} uses a $V_0$ called a ``base conformation" described as a set of atom coordinates, obtained from an auxiliary method (such as an homogeneous reconstruction or a set of atom coordinates predicted by AlphaFold \cite{alquraishi2019alphafold}). The existing reference acts as a statistical prior on the molecular volume, thereby further constraining and guiding the recovery of the conformation landscape. By contrast, \threedflex uses a reference volume $V_0$, called a ``canonical density", which is learned jointly with the conformational heterogeneity. This has the advantage of foregoing the need to split the pipeline in sequential steps, while allowing to borrow strength from the joint estimation of all parameters. 

Constraining the conformation recovery using a reference offers significant advantages for ensuring the success (and convergence) of these methods given the non-convexity of the problem. This template can be either learned (ab initio methods), or chosen from existing data (refinement methods --- more on this in Appendix \ref{appA-abinitio-vs-refinement}). The general agreement across all methods consists in tackling this hierarchically, starting with parameters which have the strongest impact on the signal, such as defocus or pose, and gradually focusing on those whose effect is more subtle, such as local deformations. As such, biasing the solution $V$ towards a reference $V_0$, such that $\Delta V(z) = 0 $ implies $V(z) = V_0$, can provide an interesting way of ensuring a more reliable and consistent --- but potentially biased --- solution. 
Depending on the optimization method used, this can in fact be critical to the success of the pipeline: Rosenbaum \textit{et al.}\cite{Rosenbaum2021} report that adopting a reference template and warm-starting their algorithm is indispensable to ensure the recovery of good conformations. However, because they fundamentally bias conformations $V(z)$ to ``hover'' around $V_0$, 
the success of such methods necessitates a reliable $V_0$. This can also incur higher computational costs, since such methods typically require running a first reconstruction method. This explains the interest for alternative, reference-free methods: three out of the six  heterogeneous methods in Table~\ref{tab:volumes} allow to recover molecular volumes without any  prior template. The extent to which these reference-free methods are likely to succeed on real-images still remains to be characterized.



\subsubsection{Defining the input domain: discrete or continuous}
The volume $V(z)$ represents a scalar 3D field (electrostatic potential, or its Fourier transform) and is defined as a function from the input domain $\Omega\subset\mathbb{R}^3$ to an output space $\mathbb{R}$ (or $\mathbb{C}$). We now describe the parametrization of $\Omega$ and distinguish two cases, depending on whether the volume is defined as a discretized or as a continuous scalar field.

\paragraph{Discretized Domain and Explicit Parametrization.} The first class of approaches models the electrostatic potential as a discrete 3D map.
In this case, the function $V(z)$ is defined on a discretized subspace (a grid) of $\mathbb{R}^3$, namely $\Omega=\{1,\ldots,D\}^3$, where $D$ represents the length of the 3D voxel grid or frequency grid. $V(z)$ is explicitly parametrized by the values it takes at each location (or voxel) of $\Omega$. This choice is also called an {\it explicit} parametrization, a term that will become clear in the next paragraph. Using a vectorial formalism, the vector $V(z)$ corresponds to voxels' intensity values, with $V(z)\in \mathbb{R}^{D^3}$ or $\in \mathbb{C}^{D^3}$. In this case, the resolution of the reconstructed volume is fixed by the choice of the granularity of the grid. However, the vectorial formalism would imply that $V(z)$ becomes an infinite-dimensional vector when it is represented continuously (see next paragraph). For this reason, we prefer to use a functional formalism and define the volume $V(z)$ as a function (not as a vector), whether it is modeled as a discrete or continuous field. Discrete domains are adopted by methods like \relion-Refine3D and \relion-Class3D\cite{Scheres2012ADetermination} --- which associate voxels with corresponding intensities in Fourier space ---, and like \cryoposenet or \threedflex in image space.

\paragraph{Continuous Domain and Implicit Parametrization.} The second class of methods model the volume $V(z)$ as a continuous field, \textit{i.e.} as a function on a continuous domain ($\Omega=\mathbb{R}^3$ or $\Omega=[-0.5,0.5]^3$). The domain $\Omega$ is infinite, and one cannot explicitly maintain in memory the values that $V(z)$ takes on $\Omega$. The solution is then to adopt an explicit parametrization for $V(z)$ using parameters $\theta\in\Theta \in \mathbb{R}^p$.
Depending on whether or not these parameters have a physical meaning (\textit{e.g.} centroids of pseudo-atoms), the function $V(z)$ can be encoded:
\vspace{1.2mm}
\begin{description}[noitemsep, leftmargin=0cm, topsep=0em,labelindent=0cm]
\item[(i) Using Neural Networks.] Some methods use neural networks to represent $V(z)$ as a (real or complex) function of a 3D position vector. The parametrization is called ``implicit" because the values of $V(z)$ are not stored in memory; instead, the practitioner can ``query" the neural network by inputing any location $x\in \mathbb{R}^3$ and receiving a value for $V(z)$ at $x$. In this case, the parameters $\theta$ --- \textit{i.e.} the weights of the neural network --- do not have a physical meaning. Examples of this approach include \cryodrgn and \cryoai, both operating in Fourier space using a reference-free volume encoding.
\vspace{1.2mm}

\item [(ii) Using Gaussian Mixtures.] Other approaches constrains the volume $V(z)$ by modeling the source of the electrostatic potential: its individual atoms or pseudo-atoms. Indeed, at a granular level, the molecular volume can be approximated by a mixture of $N$ Gaussian functions (called scattering form factors\cite{Kohl}) of the form:
\begin{equation}\label{eq:mixture}
    V_z(x) = \sum_{j=1}^N A_j \exp \left( - \frac{||c_j - x ||^2}{2\sigma_j^2}\right),
\end{equation}
where $x\in\mathbb{R}^3$ represents a 3D position, and $c_j \in \mathbb{R}^3$ are the 3D coordinates of the $N$ individual atoms or pseudo-atoms. The parameters $A_j \in \mathbb{R}$ and $\sigma^2_j \in \mathbb{R}$ describe how each (pseudo-)atom contributes to the electrostatic potential. In practice, these approaches always implement conformational heterogeneity, and do so through a continuous conformation variable $z\in \mathbb{R}^L$ that passes through a neural network to output $c_j$, and possibly $A_j$ and $\sigma^2_j$. This approach also models $V_z$ as a continuous field, as defined by Equation~\eqref{eq:mixture}, but the parameters defining each volume ($\theta=\{c_j, A_j, \sigma^2_j)\}_{j=1,...,J}$) now have a physical meaning. 
Among this general class of methods, works differ in whether $A_j, \sigma_j^2$ are assumed to be known, and in the interpretation given to the variable $c_j$. \egmm use a conformation variable $z$ that encodes the coefficients $c_j, A_j,\sigma_j^2$ and defines the $c_j$ as coordinates of ``coarse grained atoms" (reference-free). \cryofold assumes $A_j = A$ and $\sigma_j = \sigma$ known and fixed while using $c_j$ to represent ``groups of atoms''. \cryodeepmind also assumes $A_j = A$ and $\sigma_j = \sigma$ known and fixed, models the $c_j$ as the coordinates of the atoms, and uses the conformation variable $z$ to encode heterogeneous deviations $\Delta c_j.$
\end{description}

\paragraph{Discretized and Continuous Domains: Pros, Cons and Discussion.} Contrary to the discretized domains, approaches using continuous domains potentially allow to achieve sharper, enhanced resolutions (within the Nyquist limit), as any coordinate of $\mathbb{R}^3$ can be fed to $V(z)$. Moreover, within continuous approaches, pseudo-atomic methods effectively add constraints to $V(z)$ by modeling it as a mixture of Gaussians, and even more so when assuming a reference conformation $V_0$. The increasing availability of folded protein shapes --- traditionally from the Protein Data Bank\cite{ROSE2021166704} and more recently through the advent of AlphaFold\cite{alquraishi2019alphafold} --- have indeed enabled access to relatively reliable atom coordinates of reference conformations $V_0$, that can enrich the recovery of the molecular volume. We also note that reference-based representation such as that proposed in  \cryodeepmind and \cryofold are more amenable to the inclusion of molecular dynamics information to the volume reconstruction process. 

\section{Inference}\label{sec2-inference}

We now turn to the description of the inference methods used in deep generative modeling for cryo-EM reconstruction. These methods recover the volume $V$ by finding optimal parameters $\theta$, conformation variables $z_i$ and nuisance variables $(\text{PSF}_i, t_i, R_i)$ of the generative model (Eq \ref{eq:image}). In this section, $\theta$ collectively denotes the parameters that describe the conformational landscape as a function of $z$, and the parameters of the function $V_{z}:x \to \Omega$ that associates a position $x$ to an output intensity. We refer to the conformation variable and poses jointly as the ``hidden variables'' and denote them as $H_i = (z_i, \text{PSF}_i, R_i, t_i)$.
For the sake of concision, this section focuses on general inference methods, and we reserve a description of their variations for Appendix~\ref{subsec:variations_EM}. 

\paragraph{Setting Up the Inference Problem: Observed Likelihood vs Full Likelihood} 
In the context of deep generative modelling for cryoEM, the cornerstone of inference is  simply the observed  likelihood $p_\theta(x) = p(x | \theta)$ associated with each image $x$. This likelihood is computed from the generative model in Equation~\eqref{eq:image} (or its Fourier counterpart -\eqref{eq:fourier}), which we seek to maximize as a function of $\theta$. However, the generative model depends on hidden variables $H_i=(\text{PSF}_i, t_i, R_i, z_i)$. In most cases, the optimization of the full likelihood of each observation $p(x_i, h_i, \theta)$ would be quite simple, if only the $H_i$ were observed. Thus, given $n$ observed images $x_1, ..., x_n$, one solution could be to jointly recover the parameters $\theta$ and hidden variables $H$ (considered here as fixed quantities, as opposed to random variables) that maximize the log-likelihood $\ell(X, \theta)= \sum_{i=1}^n\log(p_\theta(x_i, h_i))$. Mathematically, this requires solving the following optimization problem:
\begin{align}\label{eq:loglike}
\theta^{*}, H^{*} = \text{argmax}_{\theta, H}  \sum_{i=1}^n \log(p_\theta(x_i, h_i))
\end{align}

It is in fact a classical exercise in statistics to show that in this case, as the number of estimated variables grows with the number of data points, the estimate of $\theta$ is no longer guaranteed to converge to the real underlying value as $n$ goes to infinity ($\lim_{n\to \infty} \mathbb{E}[\theta^{*}] \neq \theta^{\text{true}} $). 
We thus have to resort to strategies that treat hidden variables as random variables, and that fit the parameters $\theta$ based on the ``observed likelihood'' $L(X, \theta)$. In this case, the objective becomes:
\begin{align}\label{eq:loglike}
\theta^* = \text{argmax}_{\theta} L(X, \theta) \quad \text{ where } \quad  
L(X, \theta)= \sum_{i=1}^n \log p(x_i| \theta) 
= \sum_{i=1}^n \log \int_{h_i} p(x_i, h_i|\theta) d\mu(h_i) 
\end{align}
where $d\mu(h)= p(h) dh$ is the probability measure associated to the hidden variables $H$.
 However, this marginal likelihood requires an integral over all possible values of $H_i$. This quantity is difficult to compute directly, or in statistical terminology, ``intractable''. Consequently, the crux  of the optimization pipeline is to find a way to effectively approximate it.

\begin{figure}[h!]
\centering
    \includegraphics[width=.99\textwidth]{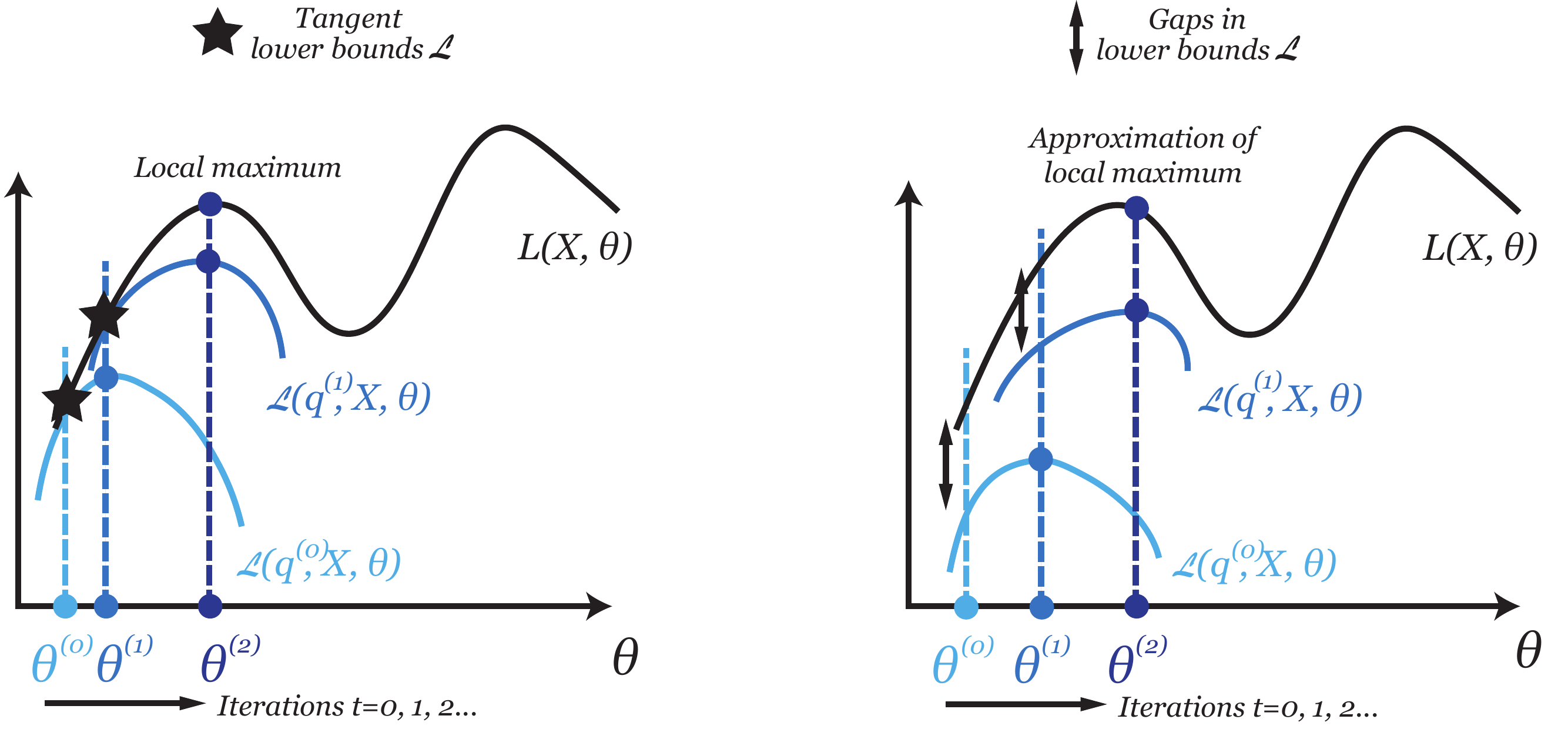}
    \caption{Maximization of the log-likelihood $\theta \rightarrow L(X, \theta)$ in $\theta$ by maximizations of a series of lower bounds: $\mathcal{L}(q^{(0)}, X, \theta)$, $ \mathcal{L}(q^{(1)}, X, \theta)$, etc. The $\theta^{(t)}$s across iterations $t=0, 1, 2, ...$ are represented by colored dots and correspond to successive maxima of the lower bounds. Left: The lower bounds are tangent to $\theta \rightarrow L(X,  \theta)$, which is realized when $q$ is the posterior of the hidden variables. Right: The lower bounds are not tangent to $\theta \rightarrow L(X,  \theta)$, but show a ``gap" that corresponds to the KL divergence between $q$ and the posterior of the hidden variables, see Equation~\eqref{eq:KL}.}
    \label{fig:lb}
\end{figure}

\subsection{Unifying Inference Methods}

Since the observed log-likelihood $L(X,\theta)$ in the objectives of Equations~\eqref{eq:loglike}-\eqref{eq:logpost} is intractable, optimization is usually performed by targeting a proxy for $L(X, \theta)$, called the \acrfull{elbo}.
For the sake of clarity and concision, we highlight here the common statistical thread of cryo-EM reconstruction methods leveraging deep generative modeling, that all use an ELBO-based optimization and refer the reader to Appendix~\ref{appC-elbo} for further discussion on their variations.

\paragraph{Evidence Lower Bound (ELBO)} 
The trick behind the  \acrfull{elbo}  consists in proposing a series of distributions $q^{(0)}, ...,q^{(t)}$ for the hidden variables $H$, and maximizing a series of ``easily'' computable lower-bounds $\mathcal{L}(q^{(0)}, X, \theta), ...,  \mathcal{L}(q^{(t)}, X, \theta)$ for $L(X,\theta)$ in an iterative fashion --- see Figure~\ref{fig:lb}. By iteratively maximizing these lower bounds with respect to $\theta$, the true likelihood $L(X, \theta)$ also increases.
The hope is that the value of $\theta$ obtained through their maximization 
will be close to the value realizing the maximum of $L(X, \theta)$, if the lower bounds are tight enough --- \textit{i.e.} for small ``gaps" in Figure~\ref{fig:lb}.

The lower bounds $\mathcal{L}(q, X, \theta)$ are found by showing that, for any probability distribution $q_i$ on the variables $h_i$, the observed log-likelihood can be written as the sum of two terms (derivations provided in Appendix~\ref{appC-elbo}):
\begin{align}\label{eq:KL}
     L(X, \theta)  =  \mathcal{L}(q, X, \theta) + \sum_{i=1}^n \text{KL}(q_i(h_i) \parallel p_{\theta}(h_i|x_i))
     = \sum_{i=1}^n \Big[ \mathcal{L}_i(q_i, x_i, \theta) +  \text{KL}(q_i(h_i) \parallel p_{\theta}(h_i|x_i)) \Big]
\end{align}
where
$KL$ is the \acrfull{kl} defined as $KL(q \parallel p)=\int  q(x) \log \frac{q(x)}{p(x)} dx$, and the terms $\mathcal{L}_i$ write:
\begin{equation}\label{eq:elbo}
\mathcal{L}_i(q_i,x_i, \theta) 
 =  \int_{h_i} q_i(h_i) \log p_\theta(x_i|h_i) dh_i - \text{KL} \left( q_i(h_i) \parallel p_\theta(h_i) \right). 
\end{equation}

The divergences $\text{KL}(q(h_i) \parallel p_\theta(h_i|x_i))$ in Equation~\eqref{eq:KL} are always non-negative. Thus, for any joint distribution $q=\{q_i\}_{i=1\cdot n}$, the function $\mathcal{L}(q,X, \theta)$ provides a valid lower-bound to $L(X, \theta)$ (see Figure~\ref{fig:lb}), called the Evidence Lower Bound (ELBO):
$$ \forall q,\forall \theta, \quad  \mathcal{L}(q,X, \theta) \leq L(X, \theta). $$

The lower-bounds $\mathcal{L}(q^{(t)},X, \theta)$ are proxies for $L(X, \theta)$, that  - in contrast to $L(X, \theta)$ - can be computed and maximized in $\theta$.

\paragraph{Inference Methods Based on an ELBO.}  While the ELBO holds for any $q$, some choices are more judicious than others. In fact, the goal is to select an optimal $q$, such that the gap between $\mathcal{L}(q,X, \theta) \leq L(X, \theta)$ is small: this will insure that the maximization of $\mathcal{L}(q,X, \theta)$ with respect to $\theta$ yields estimates $\theta^*$ that are also appropriate (and close to the true optimum ${\theta}^{\text{true}}$) for maximizing $L(X, \theta)$ --- see Figure~\ref{fig:lb} (right). Inference methods in cryo-EM subsequently differ in the choices of the distributions $q_i^{(t)}$ for each $i$ and at each iteration $t$, thereby yielding different lower bounds $\mathcal{L}(q,X, \theta)$: 

\begin{description}[noitemsep, leftmargin=0cm, topsep=0.05cm,labelindent=0cm]
\item [\bf (i) Using the posteriors given current parameters (EM algorithm):] Computing the posteriors $p_{\theta}(h_i|x_i)$ using the current estimated value $\theta^{(t)}$ of $\theta$ allows choosing $q_i^{(t)}(h_i) = p_{\theta^{(t)}}(h_i|x_i)$ for each $i$ at iteration $t$ --- see Figure~\ref{fig:posterior_distributions}. The inequality: 
    $$\mathcal{L}_i(p_{\theta^{(t)}}(h_i|x_i), X, \theta) \leq L_i(X, \theta),$$
becomes an equality for $\theta^{(t)} = \theta$. This makes the lower-bound $ \mathcal{L}(q, X, \theta)$ tangent to $L(X, \theta)$ at $\theta = \theta^{(t)}$: progressively \\
maximizing $\mathcal{L}(q, X, \theta)$ with respect to $\theta$ will induce convergence to a local maximum of $L(X, \theta)$ in $\theta$, as seen in Figure~\ref{fig:lb} (left). This is the strategy adopted by \acrfull{em} algorithm (more details in Appendix~\ref{appC-elbo}). The EM is an iterative algorithm which consists of two steps. In the first step (called the expectation step), given current parameters values $\theta^{(t)}$, we compute the posterior $q_i^{(t)}(h_i) = p_{\theta^{(t)}}(h_i|x_i)$  to plug into our ELBO. In the second (the maximization step), $\theta^{(t+1)}$ is taken to be the value of $\theta$ that maximizes the ELBO. This sequence of two steps is usually repeated until convergence. Cryo-EM methods adopting this approach are given in the first column of Table~\ref{tab:inference}. As explained in Appendix~\ref{appC-elbo}, while the EM algorithm does not have any convergence guarantees, it nonetheless guarantees to increase the likehood at each step.

\item [(ii) Approximating the posteriors given current parameters (Variational EM algorithm):] In certain cases, the choice of $q_i^{(t)}(h_i)$ as the posterior $p_{\theta^{(t)}}(h_i|x_i)$ is neither computationally attractive nor feasible. In this case, we might prefer approximating each posterior by finding its ``best approximation'' $q_i^*$ within a family of functions called variational family $\mathcal{Q}$. Cryo-EM reconstruction methods consider two choices that include approximating the posteriors by {\it (i) their ``mode", i.e. the value $\hat h_i$ of $h_i$ that maximizes $p_{\hat{\theta}^{(t)}}(x, h)$}. In this case, each $q_i$ effectively becomes a Dirac distribution at $\hat h_i$; or {\it (ii) or a general distribution $q_i$ within a family $\mathcal{Q}$}: $q_i$ is for example a Gaussian distribution -- see Figure~\ref{fig:posterior_distributions}. Cryo-EM methods adopting this approach are given in the last two columns of Table~\ref{tab:inference}, within the ``non-amortized" subcolumns.

\end{description}

\paragraph{Exact or Approximate Posteriors: Pros, Cons, Discussion.} The EM algorithm, that uses exact posteriors, holds several advantages: it is simple and stable, since all updates can only improve the observed log-likelihood. However, it is also potentially slow: the rate of convergence is known to be linear with rate proportional to the fraction of information about $\theta$ in $L(\theta, X)$ \cite{Dempster1977MaximumAlgorithm}. Variational EM algorithms can be faster; yet they potentially loose in accuracy as their ELBOs do not provide tight lower-bounds to the log-likehood $L(X, \theta)$ (Figure~\ref{fig:lb}, right). As a result, we do not have any guarantee that they converge to an (even local) maximum of $L(X, \theta)$.

\subsection{Introducing Amortized Inference}

\begin{figure}[h!]
\centering
    \includegraphics[width=0.9\textwidth]{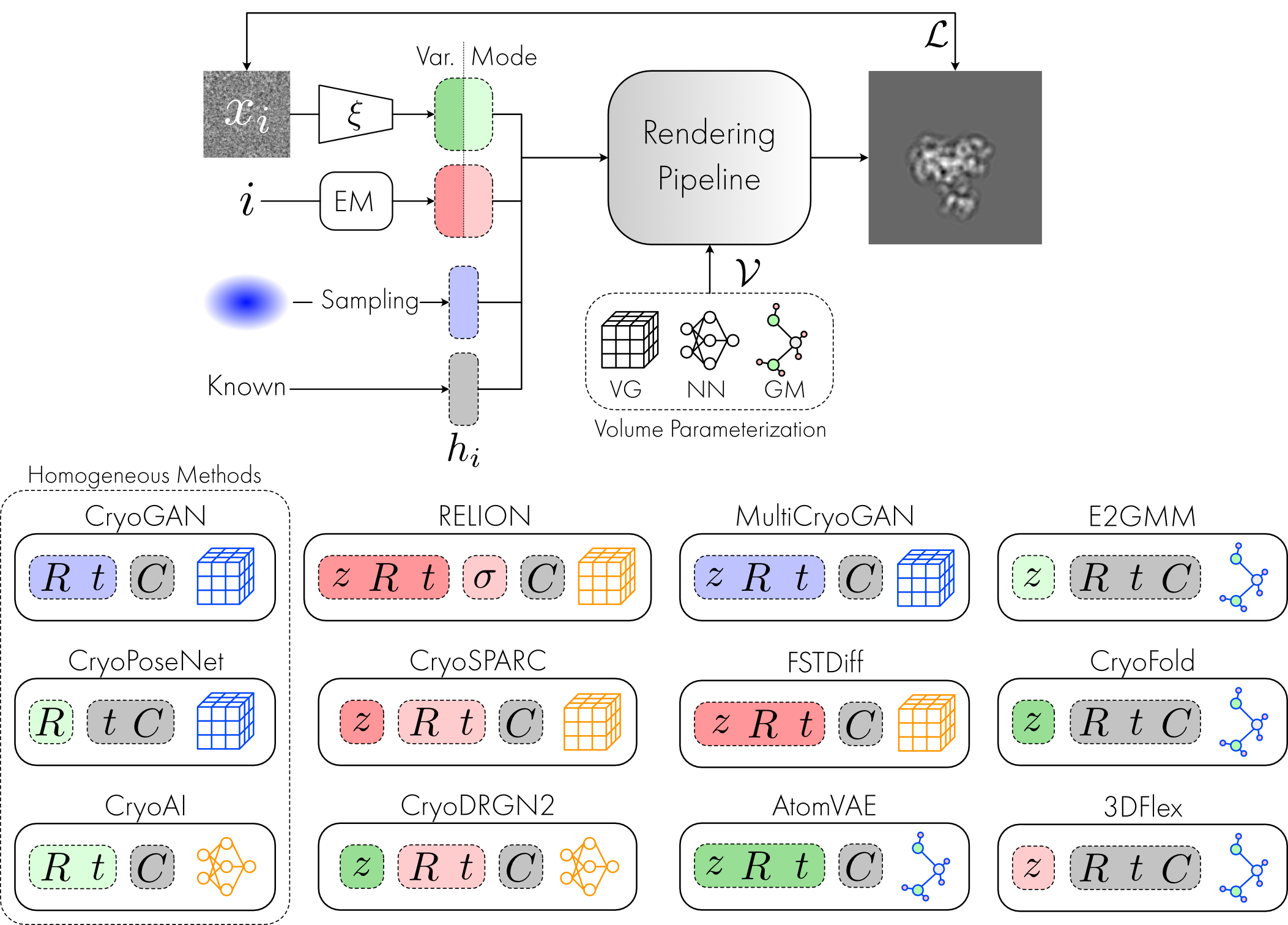}
    \caption{\textbf{Comparison of generative reconstruction methods.} Volume represented in image space (resp. Fourier space) are outlined in blue (resp. orange). CryoDRGN\cite{zhong2021cryodrgn} is similar to cryoDRGN2\cite{zhong2021cryodrgn2}, except rotations and translations are given by an upstream homogeneous reconstruction. CryoVAEGAN\cite{MiolanePoitevin2020CVPR} does not explicitly store a representation of the volume and therefore does not appear in the figure. VG = Voxel Grid ; NN = Neural Network ; GM = Gaussian Mixture.}
    \label{fig:vae}
\end{figure}


While potentially more computationally attractive than the original EM algorithm, Variational EM requires solving $n$ optimization problems to find an approximate posterior $q_i$ 
for each image $i$ in $1, ..., n$. This is computationally expensive, as the number of $q_i$ to estimate increases as the number of images $n$ increases. 
Consequently, recent methods have resorted to using an additional approximation called \acrfull{ai}, which collapses the $n$ optimizations problems into one. Instead of finding the best corresponding  $q^*_i$ for each $i$
, AI optimizes the parameters $\xi$ of a function $\text{Enc}_\xi$ that predicts the parameters of the distribution $q_i^*(h_i)$ when given $x_i$ as input, i.e.: $\text{Enc}_\xi(x_i) \simeq ( \mathbb{E}[h_i], \text{Var}[h_i]))$, where, in this example,  the variational family $\mathcal{Q}$ is chosen to be the set of Gaussian distributions.
In other words,instead of solving $n$ separate problems, Amortized Inference predicts the parameters of the posterior of image $i$ using the observed image as input. The function $\text{Enc}_{\xi}$ is traditionally called an encoder. More details --- including a description of updates performed in AI --- can be found in Appendix~\ref{appC-elbo}. Cryo-EM methods using this approach are shown in the ``amortized" subcolumns of Table~\ref{tab:inference}.

\paragraph{Implementation of Amortized Inference with Variational Autoencoders} In cryo-EM reconstruction, amortized inference is deployed in the context of variational autoencoders, denoted VAEs. VAEs are deep architectures that model the parameters of the variational family $\text{Enc}_\xi$ described above as a neural network with weights $\xi$ --- therefore leveraging the expressivity of this class of functions to get an optimal (amortized) variational approximation. The entire VAE pipeline thus consists of two steps: an encoder, which is simply a neural network with weights $\xi$ corresponding to the function $\text{Enc}_\xi$ described above, and a decoder, which allows to create ``mock samples'' that will then be compared with the observed ones based the generative model with parameter $\theta$ chosen in Section~\ref{sec1-generative_models_and_lvms}. Here, the decoder is almost entirely dictated by the process described in Equation\ref{eq:image} and goes beyond the conventional fully connected networks or convolutional neural networks used in image processing.

The learnable parameters $\xi$ and $\theta$ of the encoder and the decoder are fitted through stochastic gradient descent via backpropagation of the  ELBO  through the neural network. Compared to traditional cryo-EM reconstruction methods leveraging the EM algorithm, variational Autoencoders can be interpreted as extending the E-step (encoder) and the M-step (decoder) of the EM algorithm. The VAE architectures of cryo-EM reconstruction methods using amortized inference are given in Figure~\ref{fig:vae}. In this review, we have also included for comparison purposes a non-variational version of this procedure (the  \threedflex auto-encoder). Here, the authors consider the hidden variables as non-random variables, but add a fix amount of gaussian noise to regularise the embeddings. While the final loss is therefore adapted, this is essentially a VAE where the variance is fixed, while only the mean is learned.  



\paragraph{Amortized Inference: Pros, Cons, Discussion} Amortized inference is faster than its non-amortized counterparts, but adds an additional error (called the amortization error). We observe that several methods use amortized inference, but often to estimate one hidden variable: e.g. only the rotation $R$ or only the conformation variable $z$. Table~\ref{tab:inference} classifies the reconstruction methods by the type of inference chosen for each variable within $h_i= (R_i, t_i, \text{PSF}_i, z_i)$ and indeed, we note that this choice does not have to be consistent across all hidden variables. Many methods ``mix and match'' inference techniques, using for example a variational EM for the hidden rotation variable $R_i$ and a VAE for the conformation variable $z_i$. 
Moreover, it becomes apparent from Table~\ref{tab:inference} that (variational) autoencoders are the most common type of approaches implemented for cryo-EM reconstruction. 

While deep generative methods for cryo-EM volume reconstruction can be unified with the framework described above (as well as with traditional Expectation Maximization approaches), we observe that each of them has its own specificities or ``implementation tricks". They differ, for example, in their choice of variational family, or loss function that adapts the ELBO to facilitate convergence of the optimization procedure, see Appendix~\ref{appC-elbo}. These testify to the difficulties encountered in training these algorithms in the context of cryo-EM images with low signal-noise ratios.

\section{Discussion}\label{sec3-discussion}

Given the wide number of options to reconstruct molecular volumes from cryo-EM images, it is natural to ask: \textit{which reconstruction method is in fact the most promising?} In this last section, we focus on the need for establishing a set of metrics and benchmark tasks that can be used to quantitatively compare the performance of these methods. Starting with a review of the tools currently available, the first take-away of this section is the urgent need for new metrics and benchmarks. The evaluation of these methods' performance is currently difficult and inherently limited. We nonetheless highlight, as a second take-away, promising features in current developments, which, in our opinion, these developments pave the way for future improvements in cryo-EM reconstruction.

\subsection{Assessing Reconstruction Performances: Need for New Metrics}

Performance metrics can be categorized in two classes: (a) those that assess a method's ability to provide good \textbf{spatial resolution} (i.e. distinguishing different atoms), and, in the case of heterogeneous methods, (b) those that assess a method's ability to provide good  \textbf{conformation resolution} (i.e. distinguishing different conformations). 

\subsubsection{Assessing Spatial Resolution} 

\paragraph{Resolution of discretized reconstructions (3D maps). } When the reconstructed volume is parametrized as an explicit 3D map, the most widespread measure used to evaluate its spatial resolution is the \acrfull{fsc}\cite{haraus1986}. As described by Singer \textit{et al.}\cite{singer2020computational}, this quantity measures the correlation over a 3D shell between two reconstructed volumes:
\begin{align}\label{eq:fsc}
    FSC_k(U,V)= \frac{\sum_{s \in S_k}U_sV_s^*}{\sqrt{\sum_{s \in S_k} |U_s|^2 \sum_{s \in S_k} |V_s|^2} }.
\end{align} 
Here, $S_k$ is the set of Fourier voxels in a spherical shell at distance $k$ from the origin, and
$U$ and $V$ are the Fourier transforms of the 3D volumes that we compare. Typically, $U$ and $V$ correspond to two independent reconstructions on separate halves of the dataset, in which case, the criterion for a method to be deemed to perform well is for the two reconstructed volumes to be similar. The method's resolution is then defined as the highest resolution for which $U$ and $V$ agree ``enough''. This is precisely what the FSC (Equation \ref{eq:fsc}) captures: the FSC is close to 1 when the two maps are close. This is usually the case for small $k$, as low-frequency signal is strong, but the FSC generally decays to zero as the signal-to-noise decreases. The result is often plotted as a curve, with axis $x=k$. The resolution of the reconstruction corresponds to the maximum value of $k$ such that $FSC_k \geq 0.143$ --- a criterion chosen to match resolution criteria used in X-ray crystallography\cite{VANHEEL2005250}. For synthetic datasets where a ground-truth volume is available, the FSC is measured between the reconstruction and the ground-truth; in which case the resolution criterion correspond to the maximum value of $k$ such that $FSC_k \geq 0.5$. 

\paragraph{Resolution of continuous reconstructions. } 
Methods that represent the volume as a continuous field are relatively new, and it might be worth reassessing appropriate metrics for evaluating spatial accuracy in this case. 
\begin{description}[noitemsep, leftmargin=0.cm, topsep=0.05cm,labelindent=0cm]
\item[(i) Implicit Parametrizations] 
While interpolation between image pixels and map voxels is necessary in the discrete case, both for projection and for backprojection, implicit representations of the volume (e.g., through an neural network) enable sampling without interpolation during training. It would be interesting to investigate whether this provides a benefit in terms of reconstruction quality. We do not expect implicit representations to suddenly \emph{unlock} information, since the information content is determined by the discrete nature of the images and their pixel size, but they might provide new actionable ways to implement prior informations about the volume, such as smoothness and stereochemistry, that would result in reconstructions of higher quality.
\item[(ii) Atomic Parametrizations] Parametrizations of the volume with atomic models represents an opportunity to revisit the notion of spatial resolution. The traditional measure of similarity between two atomic models that only differ in the cartesian coordinates $U$ and $V$ of their $N$ constituting atoms (using a consistent orientation of the molecule for $U$ and $V$) is the \acrfull{rmsd}. This quantity is defined as:
\begin{align}\label{eq:rmsd}
 RMSD(U,V) = \sqrt{\frac{1}{N}\sum_{i=1}^{N}|U_i-V_i|^2},
 \end{align}
However, it was soon recognized that this metric had a very narrow range for interpretability \cite{kufareva2011methods}: it is a global measure of similarity, which is not suited to capture the local --- but meaningful--- changes in protein structure. To complicate the matter, measuring the RMSD between two atomic models assumes that they are both in the same reference frame, which might not always be defined. 
To improve the sensitivity of the metric, atomic models are routinely reduced to features with desired properties, e.g. vectors of internal coordinates independent of the reference frames. For example, reducing the atomic model to its backbone dihedral angles or to a list of atomic contacts has been shown to yield better clustering of conformations \cite{scherer2019variational}.  The development of new established metrics to evaluate these models is thus an important avenue of development.
\end{description}

\subsubsection{Assessing Conformation Resolution: An Ill-Defined Problem} 

While not flawless, the evaluation of spatial resolution is a relatively well-characterized task. By contrast, evaluating conformation heterogeneity is a more ambiguous problem. To evaluate the quality of the reconstructions allowing continuous heterogeneity, methods such as \threedflex or \cryodrgn perform a post-hoc analysis of the recovered latent space, showing the flexible deformation that are induced by sweeping through the space of possible $z$s and visually inspecting the corresponding deformations. However, proper objective and quantitative measures of conformation heterogeneity remain to be established: there currently exists no standardized measure or gold-standard task to evaluate how well a method is able to capture it.

We could design a new metric, inspired by the high-level idea of the FSC. Using two halves of the dataset to infer two independent continuous distributions of conformations, we evaluate whether the distributions agree using a metric such as the Wasserstein distance - modulo change of coordinate system for the conformation variable $z$. In the case where a ground-truth conformation is available for each image (e.g. in simulations), the inferred distribution could be compared to the true distribution. 
In the case of methods able to generate one 3D volume for each image in the dataset, one could consider a hierarchical clustering approach where depth in the hierarchy tree corresponds to the conformational resolution. In more concrete terms, for all resolution $k$, the FSC between each volume pairs would be measured and the resulting distance matrix used for clustering. Data points that fall within the same clusters would be indistinguishable at $k$ while images that would fall in different clusters would correspond to conformations that differ by at least $k$.
The development of such metrics is key to make sustainable advances in next-generation cryo-EM reconstructions. 

\subsection{Quantitative Comparison of Performances: Lack of Common Benchmarks} 

Beyond the need for new performance metrics that are better adapted to new advancements in the field, it is most certainly the lack of established benchmarks that, to this day, make reconstruction methods very hard to compare.
Such benchmarks are dramatically needed, as we cannot rely on statistical theory since the convergence properties of estimations relying on (amortized) variational inference are not completely characterized. In fact, the quantitative assessment of the methods' relative performance has yet to overcome three main hurdles:
\begin{description}[noitemsep, leftmargin=0cm, topsep=0.05cm,labelindent=0cm]
\item [\bf (i) Lack of benchmark datasets:] Current methods are developed and tested 
on a wide range of synthetic and experimental datasets that differ in the nature of the biomolecule being imaged, the dataset size, image size and associated resolution --- with very little overlap across methods - see Table~\ref{tab:experiments}. There is unfortunately no ``MNIST''\cite{deng2012mnist} or ``Imagenet''\cite{deng2009imagenet} for cryo-EM. Most methods resort to evaluating their performance on synthetic data,
yet no cryo-EM simulator acts as a standard to generate simulated images in a unified way. Synthetic datasets vary in the realism of the image formation model used for simulation, \textit{e.g.} in the noise model, the signal-to-noise ratio or the distribution of nuisance variables (e.g. poses). Subsequent experiments are typically performed on real ``in house'' data --- but there too, the important diversity within the characteristics of these evaluation datasets therefore makes the comparison of these methods a strenuous task.
\item[\bf (ii) Lack of benchmarking  procedures:] Reconstruction methods vary in the complexity of the task that they set out to accomplish, assuming more or less nuisance variables (such as poses or PSF) to be known -- see Table~\ref{tab:volumes} and Figure~\ref{fig:vae}. This makes it difficult to compare methods on a fair ground. We need to establish modular benchmarking procedure that  would enforce a  fair comparison of reconstruction performances, eg, testing the recovery of the pose, volume or conformations, with other nuisance variables being known and fixed.
\item [\bf (iii) Lack of benchmark codebase and infrastructure:] Finally, reconstructions methods are not necessarily publicly accessible, are implemented across different programming languages, and/or are tested on different software or hardware. Creating a codebase that re-implements these methods for a proper evaluation using a unified infrastructure would unfortunately represent a gigantic implementation effort. Currently, this lack of codebase poses a significant hurdle in the accessibility and comparison of the methods: it is currently impossible to disentangle the effect of their proposed statistical learning problem, their programming language, implementation tricks, or software infrastructure.
\end{description}


\subsection{Qualitative Comparison of Performances}

Despite the hurdles associated with performing quantitative comparisons, we propose a  qualitative evaluation of  the different methods based on both published results and our personal experience. This allows us to highlight promising directions  --- to the  least, in the authors' opinion --- for further developments. 

\subsubsection{Accuracy}

Despite encouraging accuracy results, some methods seem to face considerable challenges when applied to real cryo-EM images, as they have not been properly vetted and stress-tested in experimental conditions \cite{Ullrich2019,zhong2021exploring, Rosenbaum2021, Nashed2021endtoend} -- see  Table~\ref{tab:experiments}. We consider the lack of results on experimental data as a proxy for a limited applicability in the context of real signal-noise ratios regimes. In order to be adopted by cryo-EM practitionners, these methods will need to overcome the signal-noise regime and demonstrate the accuracy reported in the papers on a larger set of (benchmark) datasets.

Despite the difficulty of the task and lack of standardized benchmarks, recent developments in deep generative modeling have shown impressive promise in overcoming the current computational and accuracy bottlenecks in all three following directions:
\begin{description}[noitemsep, leftmargin=0cm, topsep=0.05cm,labelindent=0cm]
\item[\bf (i) Poses:] Poses are important nuisance variables that have the potential to damage the reconstruction, if incorrectly predicted. Accuracy of the predicted rotation is measured on synthetic datasets with a mean/median square error (MSE) against the corresponding ground-truth. Historically, preference was given to methods that did not use amortized inference for the rotation estimation (e.g. \cryosparc or \cryodrgn), as they  outperformed their amortized counterparts predicting rotations with an encoder (e.g. \cryoposenet and \cryoai): \cryodeepmind was for instance one of the only methods using amortized inference for the recovery of the poses, and reported difficulties in the joint training of poses and conformation --- highlighting the difficulty of accurate amortized inference in this setting. However, the accuracy gap between methods is closing: \cryoai now showcases an rotation accuracy at the same order of magnitude compared to \cryosparc and \cryodrgn on a real dataset. This was facilitated by the theoretical insights drawn from \cryoai, who show that amortized inference techniques tend to get stuck in local minima where the predicted molecule
contains unwanted planar symmetries due to their projections on a 2D surface.  The solution that they propose to alleviate this problem is to use the symmetrized loss:
$$ \ell_{\text{sym}} = \sum_{i}\min\{||{X}_i - \text{PSF}_i * (t_i \circ \Pi_{2D} \circ R_i) (V^{(i)})||^2, ||R_\pi({X}_i) - \text{PSF}_i * (t_i \circ \Pi_{2D} \circ R_i) ({R}_{\pi}(V^{(i)}))||^2    \}. $$
where $R_\pi$ is a rotation with angle $\pi$.
This has recently opened the door to significant gains in accuracy in the prediction of the poses, allowing for the first time pose estimation to be done through amortized inference. 
We anticipate that it is through such developments and theoretical insights that reconstruction algorithms will be able to fully leverage amortized inference for rotation prediction, hereby providing significant speed-ups.
\item[\bf (ii) Volumes:] Methods based on (pseudo-)atomic volume parametrizations - \egmm, \cryofold and \cryodeepmind - do not compare themselves to their counterparts, probably due to the fact that they were published concurrently in 2021 and/or do not use the same definition of ``pseudo-atoms" that are respectively: means of 3D Gaussian distributions, residues or actual atoms. As a consequence, we do not comment on them. For methods generating 3D maps, it has been reported by Punjani \& Fleet \cite{Punjani2021} that amortized inference can translate into resolution loss. Yet, recent methods such as \cryodrgn and \cryoai publish examples of reconstructed volumes as 3D maps whose resolution is visually comparable to the ones obtained by \cryosparc, on down-sampled imaged. Our opinion is that amortized methods can reach near-atomic resolution reconstructions, but this has not been demonstrated yet. If so, we expect them to replace traditional reconstruction pipelines in the long run, since they offer the promise to be significantly faster and to tackle much larger datasets - a desired feature to enable sufficient sampling of the conformational landscape.
\item[\bf (iii) Conformations:] Methods based on pseudo-atomic volumes parametrizations do not provide examples of conformation trajectories that allow us to compare them. For methods generating 3D maps, \cryodrgn and \threedflex seem to be some of the most promising approaches, as they seem to allow greater resolutions in the recovered trajectories, based on our personal visual assessment of the examples of conformation trajectories shown in the corresponding papers. This remains to be confirmed by a quantitative assessment over a larger number of conformation trajectories.
\end{description}

\subsubsection{Reproducibility}

Adoption of these methods by practitioners will require their reproducibility, or robustness to different initializations, implementation tricks or choice of hyperparameters:
\begin{description}[noitemsep, leftmargin=0cm, topsep=0.05cm,labelindent=0cm]
\item[\bf (i) Initialization:] The non-convex nature of the problem puts it at very high-risk of being non robust and sensitive to initialization --- this is a phenomenon sometimes referred to as ``Einstein from noise''\cite{henderson2013avoiding} (also described in Singer \textit{et al.} \cite{singer2020computational}).
Luckily, most of the current methods show encouraging signs of robustness to perturbations. In our experience, \cryodrgn seems consistent for different random initializations of the neural model when fixing the poses: the conformation space does not seem to be vastly affected.
This robustness can however be challenged by extremely low signal and/or heterogeneous datasets, in which case certain conformations can go missing. 
\item[\bf (ii) Tricks:] The inference methods presented often make use of additional implementation tricks (e.g. warm-starting with a known conformation), and specific regularization schemes: e.g. \cryodeepmind suggests starting with an initial phase of pose-only training, which, once realized, ensures that the further joint learning of poses and volume is successful. Both \cryodeepmind and \cryofold regularize the recovered structure by penalizing bond lengths, but the impact of the regularization yet remains to be properly characterized, and in particular, its potential to frustrate the optimization landscape. The importance of tricks and regularizations, and combinations thereof, is still ill-understood and would require an in-depth analysis, as it
hints towards a difficult optimization landscape for this method, and its sensitivity to initial conditions.
\item[\bf (iii) Hyperparameters:] Choosing hyperparameters such as the dimension of the latent space in algorithms such as \cryodrgn induces more or less regularization: too small and it regularizes the model too much; too large leads to underfitting of the 3D model. \cryodrgn usually sets it to $d=8$, but, given how heterogeneity arises, this is necessarily molecule dependent. The field will need --- to the least--- rule-of-thumb guidelines on how to choose these hyperparameters if these methods are to be adopted by practitioners.
\end{description}

The robustness of these new methods needs to be confirmed on a wider set of datasets, including datasets with high levels of noise. The fact that they rely on user-defined implementation tricks and hyperparameters might not be an obstacle to their adoption, as conventional methods such as \relion or \cryosparc also do.



\subsubsection{Efficiency}

Our last axis of comparison is computational efficiency: both in time and memory requirements. First, if we take the size of the datasets used in experiments as a proxy for efficiency, then \threedflex, \cryoai, \egmm, and \cryodrgn seem to be able to process remarkable amounts of information. Additionally, we offer our own practical experience by-way of rule of thumb.
With datasets typically of more than 100GB, training times can take up to 10 hours (including the required pre-processing steps) for methods like \relion --- that is, for a run that has little hyperparameter tuning compared to alternative deep learning methods. Newer methods like \cryodrgn hold great promise in terms of reconstruction: however, such sophisticated methods can further benefit from gains in efficiency, both from the computational side and in terms of memory requirements. Efficient updates of a model's parameters can thus be seen as a current computational bottleneck and offers an interesting avenue for future research.

\section{Conclusion}

This review provides a critical comparison of recent cryo-EM reconstruction methods that are based on deep generative modeling, focusing on explaining their relative advantages or drawbacks. We have unified, compared and contrasted existing methods through their parametrization of the volume, as well as through the optimization procedure chosen to recover this volume and associated hidden variables. While the use of amortized inference is crucial to make inference tractable in this high-data, high-dimensional setting, there seems to be much room for improvement and research on methods allowing both faster and better inference.
On a practical side, we note that recent methods suffer from a lack of benchmarks which severely impedes their comparison and development. From our practical experience, beyond a necessity for benchmark datasets, we also highlight a severe need for the development of a diagnostic toolbox tailored to the analysis of cryo-EM data. Current methods rely on a set of choices and hyperparameters that raise a number of questions for the practitioner: have I chosen my hyperparameters adequately? Is this choice going to impact the accuracy of the recovery? Is there any physical or biological meaning or interpretation in the distance between the latent space of conformation variables? How does error on pose or PSF affect the rest of the volume recovery process? There is therefore a pressing need for more in-depth and systematic quantitative comparison of these methods.




\printglossary[type=\acronymtype]

\printnoidxglossary

\appendix

\section*{Appendix A: Ab-initio vs Refinement Reconstructions}\label{appA-abinitio-vs-refinement}
\renewcommand{\thesubsection}{\alph{section}.}
  \setcounter{subsection}{0}

Section~\ref{sec1-generative_models_and_lvms} established how cryo-EM methods target either homogenenous or heterogeneous reconstruction, and refined these categories by drawing a distinction between methods that are reference-based, and those that are reference-free --- that is, those that do not try to parametrize heterogeneity as "small" deviations from a reference shape. Here, we introduce an additional nuance between these methods, and differentiate between methods that operate \textit{ab-initio} (ie. from scratch) , or \textit{by refinement} (ie, "warm-starting " the algorithm by using an existing and external volume estimate, such as for example a low resolution estimate). While, for the sake of clarity and conciseness, we did not elaborate on this distinction in the main text, the latter introduces an other (independent) axis of variation in our comparison of existing methods. In fact, methods can target reference-free heterogeneous reconstruction whilst warm-starting their algorithm close to a known solution (as is done in the heterogeneous version of \relion), or vice versa, prefer to opt for reference-based heterogeneity, with no prior knowledge of what the reference should look like (e.g. \threedflex).

This distinction can be understood as a way of infusing further knowledge into the recovery. \textit{From a statistical and compute science perspective}, choosing to operate by refinement depends on whether or not we have a reasonable prior for the shape (refinement), or if we'd rather not bias the algorithm with external inputs (ab initio). In this high-dimensional (and highly non-convex) setting, imposing a specific starting point for the algorithm is in fact likely to introduce an additional bias to the solution --- thereby potentially allowing it to recover more easily interesting structures. Conversely, in the absence of good guesses, ab initio methods prefer to use a random initialization. \textit{From a biological perspective}, this boils down to how much we trust potential external reconstruction methods, and we think that they could be instrumental in recovering the structure.

\begin{figure}
    \centering
    \includegraphics[width=1\textwidth]{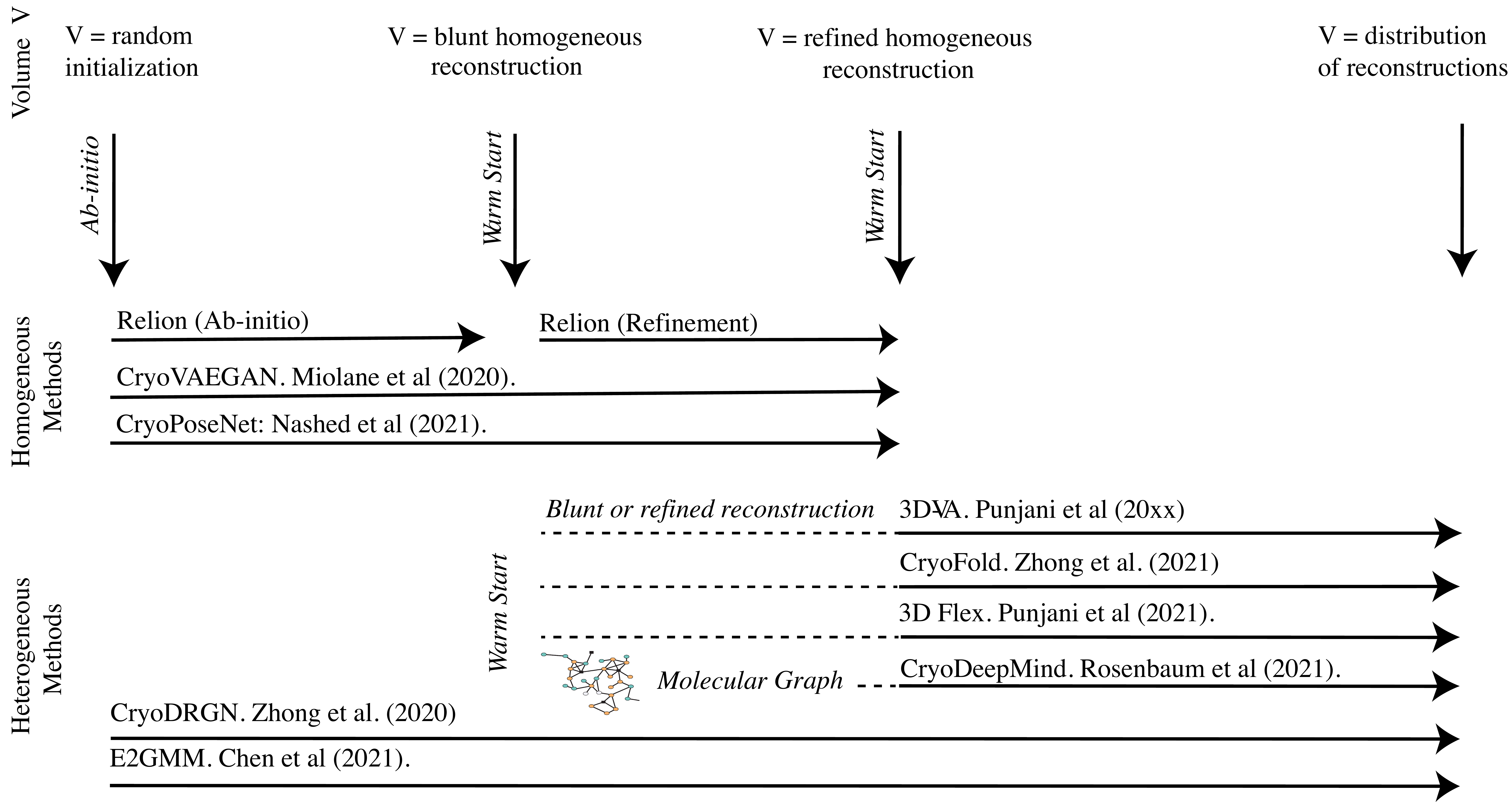}
    \caption{Works differentiated by the objective of the reconstruction (homogeneous/heterogeneous), and its initialization (ab-initio/warm-start).}
    \label{fig:cryoem_recon}
\end{figure}

\paragraph{Ab-initio vs refinement: Pros, Cons, Discussions.} Figure~\ref{fig:cryoem_recon} summarizes the methods performing ab-initio reconstruction versus refinement. While ``bias-free'', ab-initio methods can be challenging to fit: due to the highly non-convex, non-linear nature of the problem, these methods are at higher risk of  inconsistencies, especially when targeting an heterogeneous reconstruction. In fact, solutions initialised at various random points might not necessarily converge to same point, and/or get stuck in local minima. By contrast, warm-starting the problem can yield faster, and more consistent convergence towards the solution. Warm-starting the solution might also increase consistency across recovered conformation, thereby preventing aberrations. 
Homogeneous reconstruction approaches can also benefit from warm-starting strategies: in RELION \cite{Scheres2012RELION} for instance, the hypothesised template is converted into Fourier coefficients, that are used in the initialisation part of the inference process. Refinement for homogeneous reconstruction holds several advantages, especially if the homogeneous reconstruction is either slow or very sensitive to initialization. 

Yet, refinement also holds several disadavantages. This approach requires a two-step procedure and another fragmentation of the estimation pipeline, as it relies on a prior estimation of an initial "blunt" value by another pipeline. Moreover, as highlighted in the review of Bendory et al \cite{bendory2020single}, it also exposes this solution to model bias: the structure recovered will be biased towards our initial prior --- potentially hindering our ability to detect flaws in this original template or discovering new conformations. Finally, while the literature on the sensitivity of the methods to an erroneous warm-starting is also nonexistent, the impact of a wrong initialization could also either severely hinder the performance of the algorithm, or push the algorithm to hallucinate nonexistent solutions (allowing us to recover ``Einstein from noise'' \cite{henderson2013avoiding}--- see discussion in part \ref{sec3-discussion}). 

\section*{Appendix B: Differentiability of the Generative Models}\label{appB-operators}

As described in the main text, an alternative (and often simpler) generative model can also be formulated via the Fourier equivalent of Equation~\eqref{eq:image}. 

 \paragraph{Fourier-Slice Theorem.}
In Fourier Space, the convolution $*$ of the projected 2D volume by the PSF conveniently becomes an element-wise matrix multiplication $\odot$ between the 2D Fourier transform of the projected image and that of the PSF. The latter is better known as the Contrast Transfer Function (\acrshort{ctf}), with $\text{CTF}_i = \mathcal{F}(\text{PSF}_i)$, so that:
\begin{equation}\label{eq:fourier}
\tilde{X}_{i} = \text{CTF}_{i} \odot (P_i^{(R_i,t_i)} \circ \tilde{V}^{(i)}) + N_{i}, \qquad \text{with} \quad i=1 \cdots n.
\end{equation}
\noindent Here, $\tilde{X}_{i}$ is the 2D-Fourier transform of the observed image $X_i$ ($\tilde{X}_{i}=\mathcal{F}(X_i)$), and $\tilde{V}^{(i)}$ is the 3D Fourier transform of the molecule (i.e, the volume) $V^{(i)}$, with $\tilde{V}_{i}=\mathcal{F}(V_i)$. In Equation~\eqref{eq:fourier}, the operator $P_i^{(R_i,t_i)}$ first extracts out of $\tilde{V}^{(i)}$ a slice oriented by the rotation $R_i$ and then applies a phase shift in Fourier space accounting for the 2D translation $t_i$.  Finally, $N_{i}$ represents the noise. We note that using the operator $P_i$ circumvents the 3D-to-2D projection step and is therefore more computationally efficient.


\paragraph{Efficiency and differentiability of the different generative models} The methods described in this review  (Eq~\ref{eq:image} and \ref{eq:fourier}) leverage the ability to do inference with gradient-based optimization techniques. This requires the generative models to be differentiable --- that is, to allow the differentiation of the loss associated with the output (an image) with respect to the parameters of the model through backpropagation. As explained in Section~\ref{sec1-generative_models_and_lvms}, the volume $V$ (or $\tilde{V}$ in Fourier space) can always be seen as a function from $\mathbb{R}^3$ to $\mathbb{R}$ (or $\mathbb{C}$). The parameters defining this function depend on the choice of \textit{parametrization}, as summarized in Table~\ref{tab:diff}. To make sure that the model is differentiable, we need to ensure that all the operations required in the forward model presented by Equation~\ref{eq:image} are differentiable. Table ~\ref{tab:diff} summarizes the operations required to ``rotate'', ``translate'' and ``project'' the volume $V$ and obtain the observed image. Those operations are:
\begin{itemize}[noitemsep]
    \item multiplication with a matrix $R\in\mathbb{R}^{3\times 3}$,
    \item addition of a vector $\mathbf{t} \in \mathbb{R}^3$,
    \item application of an interpolation kernel,
    \item summation (if the volume is in image space),
    \item and analytical integration.
\end{itemize}
All these operations, if they involve the ``differentiable parameters'' can be differentiated through by backpropagation. As shown in Equation~\eqref{eq:image} (resp. Equation~\eqref{eq:fourier}) the last step of the generative model is the application of the PSF (resp. CTF) which is done with a convolution (resp. element-wise multiplication) with the projected volume. This last operation is also differentiable, which in turn ensures that the whole generative model, transforming a volume $V$ into an image (potentially in Fourier space), is differentiable.

\begin{table}[h]
\centering
\begin{tabular}{c|c|c|c|c|}
\cline{2-5}
\rowcolor[HTML]{EFEFEF} 
\cellcolor[HTML]{FFFFFF}                                                                                     & \textbf{\begin{tabular}[c]{@{}c@{}}Differentiable\\ Parameters\end{tabular}} & \textbf{Domain}                                                                                                          & \textbf{\begin{tabular}[c]{@{}c@{}}Rotation\\ Translation\end{tabular}}                                   & \textbf{\begin{tabular}[c]{@{}c@{}}Projection\\ (in image space)\end{tabular}}                                                                                         \\ \hline
\multicolumn{1}{|c|}{\cellcolor[HTML]{EFEFEF}\begin{tabular}[c]{@{}c@{}}Voxel\\ Grid\end{tabular}}           & \begin{tabular}[c]{@{}c@{}}$V(x)$\\ for $x\in\{1,\ldots,D\}^3$\end{tabular}    & \begin{tabular}[c]{@{}c@{}}$V: \{1,\ldots,D\}^3$\\ With interpolation kernel,\\ $\hat{V}: \mathbb{R}^3$, see \cite{Ullrich2019}\end{tabular}     &                                                                                                           & \begin{tabular}[c]{@{}c@{}}$\sum_x \hat{V}\hat{(x)}$\\ $\hat{V}:\mathbb{R}^3\rightarrow\mathbb{R}$\end{tabular} \\ \cline{1-3} \cline{5-5} 
\multicolumn{1}{|c|}{\cellcolor[HTML]{EFEFEF}\begin{tabular}[c]{@{}c@{}}Neural\\ Network\end{tabular}}       & Weights $W$                                                                    & $V: \mathbb{R}^3$                                                                                                        & \multirow{-2}{*}{$\hat{x}=R^{-1}x-\mathbf{t}$}                                                            & $\sum_x V\hat{(x)}$                                                                                             \\ \hline
\multicolumn{1}{|c|}{\cellcolor[HTML]{EFEFEF}\begin{tabular}[c]{@{}c@{}}Mixture of\\ Gaussians\end{tabular}} & $A_j$, $c_j$, $\sigma_j$                                                       & \begin{tabular}[c]{@{}c@{}}$V: \mathbb{R}^3$\\ With analytical integration,\\ $\int V(.,z)dz: \mathbb{R}^2$\end{tabular} & \begin{tabular}[c]{@{}c@{}}$c_j\rightarrow\hat{c}_j=Rc_j+\mathbf{t}$\\ $V\rightarrow\hat{V}$\end{tabular} & \begin{tabular}[c]{@{}c@{}}$\int \hat{V}(y,z)dz$\\ for $y\in\mathbb{R}^2$\end{tabular}                        \\ \hline
\end{tabular}
\caption{Differentiable parameters and operations realized in the generative model, depending of the volume parametrization. $V$ can be equivalently replace with $\tilde{V}$ in Fourier space. Using an interpolation kernel transforms the function $V$ defined on $\{1,\ldots,D\}^3$ into $\hat{V}$ defined on $\mathbb{R}^3$.}
\label{tab:diff}
\end{table}

\section*{Appendix C: Inference and ELBO Computations}\label{appC-elbo}
\renewcommand{\thesubsection}{C.\arabic{subsection}}

This appendix provides more details on the Bayesian formulation of the cryoEM recovery problem, and on the corresponding methods. As a reminder, in the main text, we argue that the lower bounds $\mathcal{L}(q, X, \theta)$ are found by showing that, for any probability distribution $q_i$ on the variables $h_i$, the observed log-likelihood can be written as the sum of two terms:
\begin{align}\label{eq:KL2}
     L(X, \theta)  =  \mathcal{L}(q, X, \theta) + \sum_{i=1}^n \text{KL}(q_i(h_i) \parallel p_{\theta}(h_i|x_i))
     = \sum_{i=1}^n \Big[ \mathcal{L}_i(q_i, x_i, \theta) +  \text{KL}(q_i(h_i) \parallel p_{\theta}(h_i|x_i)) \Big]
\end{align}
where
$KL$ is the \acrfull{kl} defined as $KL(q \parallel p)=\int  q(x) \log \frac{q(x)}{p(x)} dx$, and the terms $\mathcal{L}_i$ write:
\begin{equation}\label{eq:elbo2}
\mathcal{L}_i(q_i,x_i, \theta) 
 =  \int_{h_i} q_i(h_i) \log p_\theta(x_i|h_i) dh_i - \text{KL} \left( q_i(h_i) \parallel p_\theta(h_i) \right). 
\end{equation}
The objective of this appendix is to derive these bounds and present the corresponding variational techniques in greater details.

\subsection{A Bayesian take on  cryo-EM}

Before we show these bounds, let us begin by a preliminary remark regarding the model that we are trying to fit. In our formulation in the main text, the parameters $\theta$ are considered as unknown, fixed values. Alternatively, one could also try and incorporate the uncertainty of these parameters by adopting a hierarchical approach, and modelling the parameters $\theta$ as random variables with prior $p(\theta)$. In this case, inference seeks to estimate $\theta$ through the maximum of its posterior distribution $p(\theta | x_1, ..., x_n)$, written in its logarithm form as:
\begin{align}\label{eq:logpost}
\log p(\theta | x_1, ..., x_n) 
\propto \log p(\theta, x_1, ..., x_n)
= \log p_\theta(x_1, ..., x_n) + \log p(\theta)
= L(X, \theta) + \log p(\theta).
\end{align}
This can increase robustness by embedding in the design of the algorithm the inherent variability and uncertainty associated with $\theta$. In fact, following Scheres \cite{Scheres2012RELION}, adding priors on $\theta$ can be viewed as a kind of regularization. Such approaches --- referred to as Maximum a Posteriori (MAP) approaches --- are chosen in Relion-Refine3D and Relion-Class3D\cite{Scheres2012ADetermination}, \cryomaxwelling, and \cryosparc.  In the original RELION\cite{Scheres2012RELION}  method for instance, the underling volume is represented through its Fourier components $V_l$, whose prior is chosen to be zero-mean Gaussian distributedwith unknown variance $\tau_l^2$.

We observe that the ML and MAP objectives in Equations~\eqref{eq:loglike}-\eqref{eq:logpost} only differ through the prior term $\log p(\theta)$, while the challenge of the optimization stems froms the intractable term $ L(X, \theta)$. Both approaches traditionally employ the same optimization techniques.

\subsection{ELBO Computations}

We provide the derivations leading to Equations \eqref{eq:KL2}-\eqref{eq:elbo2} and the definition of the ELBO as a lower bound to the log-likelihood of the generative model. We start by expressing the observed log-likelihood of the model as:
\begin{align*}
    L(X, \theta) 
    &= \sum_{i=1}^n\log p(x_i| \theta) \\
    &= \sum_{i=1}^n \left( \log p(x_i,h_i| \theta) - \log p(x_i,h_i| \theta) + \log p(x_i| \theta) \right)\\
    &= \sum_{i=1}^n\left( \log p(x_i,h_i| \theta) -  \log p(h_i|x_i, \theta) \right) \qquad \text{( Bayes rule:}\quad p(h_i| x_i, \theta) = \frac{p(x_i, h_i|\theta)}{p(x_i|\theta)}) 
\end{align*}
In what follows, we drop the index $i$ and associated sum for convenience. Consider any distribution $q(z)$. If we multiply both sides of the above equation by $q$, and integrate the $z$ out, we get:
\begin{align*}
    \int_{z} L(X, \theta)q(h)dh &= \int_h\left(
            \log p(x,h| \theta) 
                - \log p(h|x, \theta)\right)q(h)dh\\
    \implies   L(X, \theta)&= \int_h\left(
            \log p(x,h| \theta) 
                - \log p(h|x, \theta)\right)q(h)dh,
\end{align*}
since $L(X, \theta)$ does not depend on $h$. Now expanding the right hand side:
\begin{align*}
    L(X, \theta)
    &= \int_h q(h)\log p(x,h| \theta) dh 
        - \int_h q(h) \log p(h|x, \theta) dh  \\
    &= \int_h q(h)\log p(x,h| \theta) dh 
        - \int_h q(h) \log \frac{p(h|x, \theta)q(h)}{q(h)} dh  \\
    &= \int_h q(h) \log p(x,h| \theta) dh  
        - \int_h q(h) \log q(h) dh
        - \int_h q(h) \log \frac{p(h|x,\theta)}{q(h)}dh \\
    &= \int_h q(h) \log p(x,h| \theta) dh  
        - \int_h q(h) \log q(h) dh
        + KL(q(h) \parallel p(h|x,\theta))\\
    &= \mathbb{E}_q[ \log(p_{\theta}(x, h) ] 
        - \int_h q(h) \log q(h) dh
        + KL(q(h) \parallel p(h|x,\theta))  \\ 
    &= \mathbb{E}_q[ \log(p_{\theta}(x, h) ] 
        + H(q) + KL(q(h) \parallel p(h|x,\theta)),
\end{align*}
by introducing the entropy of $q$ as $H(q) = - \int_z q(z) \log q(z)dz$, and writing the KL-divergence between two distributions $p_1$ and $p_2$ as $KL(p_1||p_2) = \int p_1(x) \log ( \frac{p_1(x)}{p_2(x)})dx$.\\

In other words, for any distribution $q$ on the latent variable $H$, the observed marginal likelihood $L(X, \theta)$ is the sum of three terms: (a) the expected log-likelihood, assuming that $h$ has distribution $q$, (b) the Entropy of $q$, which can be understood of the amount of uncertainty in the estimation of $h$ and (c) the KL-divergence between $q$ and the true posterior $ p(h|x, \theta)$, which measures how different these distributions are.\\

Note that, in the previous equations, terms (a) and (b) can be computed, but term (c) --- the KL divergence between the proposed distribution $q$ for $h$ and the actual posterior distribution $p(h|x, \theta)$ is not unknown, because we do not know $p(h|x, \theta)$. However, because the KL-divergence is always non-negative, we know that:
$$ L(X, \theta) \geq   \mathbb{E}_q[ \log(p_{\theta}(x, h) ] + H(q).$$
Thus, introducing the ELBO term $\mathcal{L}(q, X, \theta) = \mathbb{E}_q[ \log(p_{\theta}(x, h) ] + H(q)$, we get:
\begin{equation}\label{eq:ell}
    L(X, \theta)
    = \mathcal{L}(q, X, \theta) 
    + KL(q(h) \parallel p(h|x,\theta)).
\end{equation}
which provides Equation~\eqref{eq:KL2} and the right hand side of Equation~\eqref{eq:elbo2}. Developping the ELBO term gives:
\begin{align*}
    \mathcal{L}(q, X, \theta)
    &=  \int_h q(h) \log p(x,h| \theta) dh  
        - \int_h q(h) \log q(h) dh \\
    &=  \int_h q(h) \log p(x| h, \theta) dh  
        +  \int_h q(h) \log p(h| \theta) dh  
        - \int_h q(h) \log q(h) dh  \\
    &=  \int_h q(h) \log p(x| h, \theta) dh  
        - KL(q(h) \parallel p(h|\theta),
\end{align*}
which gives the left side of Equation~\eqref{eq:elbo2}.

\subsection{Expectation-maximization (EM) Algorithm}\label{subsec:em}

Historically, inference in latent variable models has been achieved through the Expectation-maximization (EM) algorithm \cite{Dempster1977MaximumAlgorithm}. The EM algorithm is often introduced as a data imputation technique (see following subsection), but it can also be understood as a dual ascent algorithm --- a perspective that allows unifying most cryoEM inference methods and that we present here.

\begin{figure}
 \vspace*{-0.3in}
 \centering
    \includegraphics[width=0.4\textwidth]{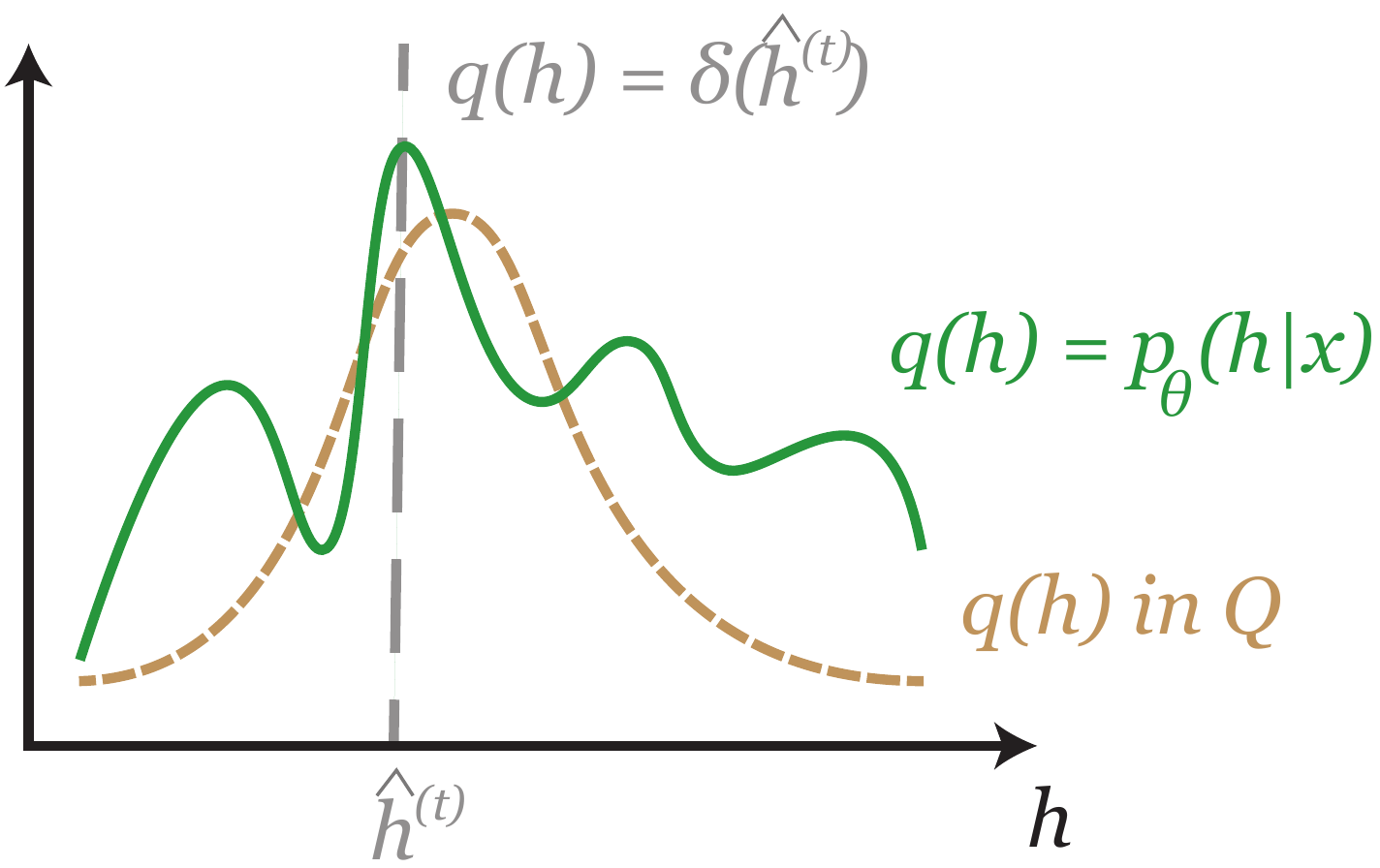}
    \caption{Possible choices for the distribution $q$ on the hidden variables $h$, where we have dropped the indices $i$ for convenience of notations. The choice of $q$ determines the lower bounds to $L(X, \theta)$ in Figure~\ref{fig:lb}. The inference methods used in cryoEM reconstruction can be contrasted by the choice of $q$: as the true posterior $p_\theta(h|x)$ represented by the green line, as the Dirac distribution $\delta(h)$ represented by the vertical dashed gray line or as a distribution within a known parameterized family $\mathcal{Q}$, such as the family of Gaussian distributions, represented by the brown dashed line.}
    \label{fig:posterior_distributions}  \vspace*{-0.18in}
\end{figure}

\paragraph{EM as a Dual Ascent Algorithm} The EM algorithm can be seen as a dual ascent procedure, \textit{i.e.} as a maximization-maximization procedure, that leverages Equation~\eqref{eq:KL2} rewritten as: $\mathcal{L}(q, X, \theta) = L(X, \theta)  -  \sum_{i=1}^n \text{KL}(q_i(h_i) \parallel p_\theta(h_i|x_i))$. Starting with an initial guess $\theta^{(0)}$ of the parameter, the EM algorithm performs two steps at each iteration $t$:
  \begin{description}[noitemsep]
  \item[(a) Inference on hidden variables (E-step)] Given the current $\theta^{(t-1)}$, maximize each term $\mathcal{L}_i(q_i, \theta^{(t-1)}) = L_i(X_i,  \theta^{(t-1)})  - KL(q_i(h_i) || p_{ \theta^{(t-1)}}(h_i,x_i))$ (by Equation \ref{eq:ell}) with respect to the distribution $q_i$ defined on the hidden variables $h_i$. Since $\theta^{(t-1)}$ is fixed, the best choice of $q$ is the one that minimizes the KL term $KL(q_i(h_i) || p_{ \theta^{(t-1)}}(h_i,x_i))$:
  \begin{equation}
       \text{For each $i = 1, ..., n$:} \quad\quad q_i^{(t)}(h_i) = \argmax_q \mathcal{L}_i(q, X, \theta^{(t-1)})= \argmin_q KL(q_i(h_i) || p_{ \theta^{(t-1)}}(h_i,x_i))  = p_{\theta^{(t-1)}}(h_i| x_i).
  \end{equation}
  
  This choice $q_i^{(t)}(h_i)=p_{\theta^{(t-1)}}(h_i| x_i)$ corresponds to the green distribution in Figure~\ref{fig:posterior_distributions} and makes the lower bound $\mathcal{L}$ tangent to $L$ at $\theta^{(t-1)}$ in Figure~\ref{fig:lb} (left).
  
    \item[(b) Maximization on the model's parameters (M-step)] Given the current $q^{(t)} = \{q_1^{(t)}, ..., q_n^{(t)}\}$, maximize $\mathcal{L}(q^{(t)}, X, \theta)$ with respect to the model parameters $\theta$:
    \begin{equation}
        \theta^{(t)} = \argmax_{\theta} \mathcal{L}(q^{(t)}, X, \theta) = \argmax _{\theta} \sum_{i=1}^n \mathcal{L}_i(q_i^{(t)}, X, \theta).
    \end{equation} 
    The argument maximum of $\mathcal{L}$ is used to update $\theta^{(t-1)}$ to $\theta^{(t)}$ in Figure~\ref{fig:lb} (left).
  \end{description}
These steps are iterated until convergence in $\theta$. In other words, one first finds the "best" lower bound to $L(X, \theta)$ given $\theta^{(t)}$ by choosing the one that is tangent to $L(X, \theta)$ at $\theta^{(t)}$. Once this bound has been established, we subsequently maximize this bound with respect to the parameter $\theta$.
We note that to alleviate potential computational bottleneck in the  M step (respectively $\mathcal{L}(q^{(t)}, X, \theta)$), the closed form solution can be replaced by simply taking a gradient step. This gives rise to ``gradient EM" methods, which we will refer to later, as all methods introduced in the next subsections can be formulated as ``gradient" methods too.

\paragraph{Computational Bottleneck: E-step.} One of the issues with the EM algorithm relies in the computation of  the posterior of the latent variable $h_i$ given $x_i$ and current estimate $\theta^{(t-1)}$:
\begin{equation}\label{eq:post}
p_{\theta^{(t-1)}}(h_i|x_i)  = \frac{p_{\theta^{(t-1)}}(x_i|h_i)p(h_i)}{p_{\theta^{(t-1)}}(x_i)} = \frac{p_{\theta^{(t-1)}}(x_i|h_i)p(h_i)}{\int_{h_i} p_{\theta^{(t-1)}}(x_i, h_i) dh_i}.
\end{equation}

Unless the posterior can be computed efficiently, this computation can be lengthy. In fact, unless the priors of the different variables are specified using conjugacy --- which would constrain the choice of prior distributions that we could consider ---, it is difficult to get closed-form updates for the posterior. Alternatively, one can evaluate $p_{\theta^{(t-1)}}(h_i|x_i)$ for each $h_i$ discretized on a grid by computing the integral in Equation~\eqref{eq:post} via a Riemann sum as in \relion or via Importance Sampling as in \cryosparc. \relion and \cryosparc use the EM algorithm with MAP estimation of $\theta$, as opposed to ML estimation. 

This discretization remains however a computationally intensive approach. Consequently, despite many advances leveraging GPU computing, the integral in Equation~\eqref{eq:post} represents the main computational bottleneck in cryo-EM reconstruction methods, and the main reason we might want to look for other, more efficient alternatives. 


\subsection{Variations on the EM algorithm}\label{subsec:variations_EM}

We present details on the variations on the EM algorithm discussed in the main text.

\subsubsection{Modal EM algorithm}\label{subsec:mem}

The computational bottleneck observed in the EM algorithm has also motivated the use of approximations in the E-step. The modal approximation of EM amounts to finding a proxy for the posterior $p_{\theta^{(t-1)}}(h_i|x_i)$ in the E-step. This approximation replaces the posterior by its "mode" $\hat{h}_i = \argmax_h p_{\theta^{(t-1)}}(h|x_i)$, i.e. the most probable $\hat h_i$ for each image $x_i$. 

\paragraph{Modal EM: The algorithm.} The E-step of the EM algorithm is replaced by its modal approximation, but the M-step is largely unchanged:
\begin{description}
  \item[(a) Inference on hidden variables $h_i$] Given current $\theta^{(t)}$, compute the modes:
  \begin{equation}
  \text{For each $i = 1, ..., n$:} \quad\quad
  \hat{h}_i^{(t)} = \argmax_h p_{\theta^{(t-1)}}(h|x_i) = \argmax_h p_{\theta^{(t-1)}}(x_i|h)p(h).
  \end{equation}
  
  This amounts to approximating the solution of the E step (which, we've shown, is the posterior of $h$ given $X$ and $\theta^{(t-1)}$) by  $q_i^{(t)}(h_i)=\delta (\hat h_i^{(t)})$, the Dirac distribution at $\hat h_i^{(t)}$, represented in gray in Figure~\ref{fig:posterior_distributions}. This choice for $q_i^{(t)}(h_i)$ creates a lower bound $\mathcal{L}$ that may not be tangent to $L$ in Figure~\ref{fig:lb} (right). Since this approximation does not require us to compute complicated posteriors, it is a simpler, but looser ELBO to the observed likelihood.
  \item[(b) Maximization on model's parameter $\theta$] The parameter $\theta$ is updated via:
  \begin{equation}\label{eq:modal-m}
      \theta^{(t)} = \argmax_\theta \sum_{i=1}^n \mathcal{L}(\delta (\hat h_i^{(t)}), \theta) =  \argmax_\theta \sum_{i=1}^n p_{\theta}(x_i, \hat h_i^{(t)}).
  \end{equation}
  The argument maximum of $\mathcal{L}$ is used to updated $\theta^{(t-1)}$ to $\theta^{(t)}$ in Figure~\ref{fig:lb} (right), just as in the EM algorithm in Figure~\ref{fig:lb} (left).
 \end{description}
 


\paragraph{Modal EM in \relion, \cryosparc, \threedflex, \cryodrgn, \cryofold} The EM algorithm with modal approximation is used in cryo-EM in \cryosparc, \threedflex, \cryodrgn and \cryofold to infer the latent variable associated with the rotation $R_i$. In other words, they estimate the best rotation $\hat R_i$ for each input image $i$, relying on a branch-and-bound optimization algorithm to address this maximization. This approach is also used in \relion to estimate the best noise variance $\hat \sigma_i$ that is associated with each image $i$.

\subsubsection{Variational EM algorithm}\label{subsec:vem}

The modal approximation in the EM algorithm speeds up the E-step; yet it has the drawback of summarizing the whole posterior distribution $p_{\theta^{(t)}}(h_i|x_i)$ by a single estimate $\hat h_i$, reducing accuracy and leading to non-tangent lower bounds $\mathcal{L}$. \acrfull{vi}, also known as Variational Bayes, has appeared in cryo-EM as a compromise between the guarantees of the EM and the efficiency of modal EM during the computation of the E-step. 

\acrfull{vi} replaces the evaluation of the posterior of the latent variables $p_{\theta^{(t-1)}}(h_i|x_i)$, by an optimizing over a different family of candidate distributions $\mathcal{Q}$, called the ``variational family''. $\mathcal{Q}$ is typically a family with a restricted form, so that the updates are easier to perform: for instance, if there is a complex dependency structure between entries in our parameter $\theta$, we might select $\mathcal{Q}$ to be a family of posteriors where the entries are independent, so that the updates are simpler to perform: this is usually the `mean field' approximation to the posterior of $\theta$. The $q$ that is chosen in Equation \ref{eq:elbo2} is the "closest" to the posterior, where ``closest" is defined in terms of the KL divergence (see Equation~\eqref{eq:vi} below). 



The variational family $\mathcal{Q} = \{ q_\eta | \eta\}$ with parameters $\eta$ is typically chosen to be a family of Gaussian distributions, in which case $\eta$ represents the mean and variance, i.e. $\mathcal{Q} = \{ q_\eta = N(\mu, \sigma) | \eta=(\mu, \sigma)\}$. Other approaches consider the variational family of Gaussian distributions with fixed isotropic variance $\sigma_0$, i.e. $\mathcal{Q} = \{q_\eta = \mathcal{N}(\eta, \sigma_0) | \eta\}$.  In  modal EM, the parametric family can be seen as a family of Dirac distributions, as in Equation~\eqref{eq:modal-m}.


\paragraph{Variational EM as a Dual Ascent Algorithm} The E-step of the EM algorithm is replaced by VI and the M-step is unchanged. The VI framework thus gives rise to a variational EM which performs the following two steps at each iteration $t$:
\begin{description}
  \item[(a) Inference on hidden variables $h_i$] The $n$ parameters $\eta_i$ corresponding to each $q_i$ are computed via:
  \begin{equation}\label{eq:vi}
      \text{For each $i = 1, ..., n$:} \quad\quad \eta_i^{(t)} = \argmin_\eta KL(q_{\eta}||p_{\theta^{(t-1)}}(h_i|x_i)
  \end{equation}
  This choice for $q$ is represented in brown in Figure~\ref{fig:posterior_distributions}. Similarly to the Modal EM, this creates a lower bound $\mathcal{L}$ that may not be tangent to $L$ in Figure~\ref{fig:lb} (right).
  \item[(b) Maximization on model's parameter $\theta$] The parameter $\theta$ is updated via:
  \begin{equation}
      \theta^{(t)} = \argmax_\theta \sum_{i=1}^n \mathcal{L}_i(q_{\eta_i}^{(t)}, \theta).
  \end{equation}
    The argument maximum of $\mathcal{L}$ is used to updated $\theta^{(t-1)}$ to $\theta^{(t)}$ in Figure~\ref{fig:lb} (right), just as in the EM and the Modal EM algorithms.
 \end{description}

Thus, the Variational Inference approximation in the E-step replaces the samplings in evaluation of the true posterior, required to compute the integral in Equation~\eqref{eq:post}, by an optimization on the fewer parameters $\eta_i$ parametrizing each approximate posterior $q_{\eta_i}$. This optimization can be efficiently conducted with (stochastic) gradient descent. In this sense, it also represents a solution to the computational bottleneck of the E-step.

\paragraph{Variational EM in \threedflex and \cryomaxwelling} Variational gradient EM is used in \threedflex to infer the rotation variable $R_i$, using the variational family of Gaussian distributions with fixed isotropic variance $\sigma_0$.
Variational EM is also used in \cryomaxwelling in the context of homogeneous reconstruction. Here, VI happens both in Step (a), to estimate the approximate posterior of the rotation $R_i$, and in Step (b) to also estimate the approximate posterior of the volume $V$, which is a parameter included in $\theta$. The approximate posterior of $V$ provides an approximate measure of uncertainty on the homogeneous reconstruction.


\paragraph{Amortized Variational EM as Dual Optimization Algorithm} Amortized Inference in the E-step of variational EM leads to an amortized variational EM. In this case, the algorithm can be written as:
\begin{description}
  \item[(a) Inference on hidden variables $h_i$] The parameters $\xi$ parametrizing the encoder is computed via:
  \begin{equation}
      \xi^{(t)} = \argmax_\xi \sum_{i=1}^n KL(q_{\text{Enc}_\xi(x_i)}||p_{\theta^{(t-1)}}(h_i|x_i)
  \end{equation}
  which generates $n$ distributions $q_{\eta_i}^{(t)}$ parametrized by $\eta_i = \text{Enc}_{\xi^{(t)}}(x_i)$.
  \item[(b) Maximization on model's parameter $\theta$] The parameter $\theta$ is updated via:
  \begin{equation}
      \theta^{(t)} = \argmax_\theta \mathcal{L}(q_{\eta_i}^{(t)}, \theta).
  \end{equation}
 \end{description}

This method has the advantage of further improving  the flexibility and efficiency of the Variational EM algorithm, by allowing the updates to be even more efficient by parametrizing the $q_i$ as a function of the input $x_i$. 

\subsection{Generative Adversarial Networks}\label{subsec:gans}

Cryo-EM reconstruction methods from Subsections~\ref{subsec:mem}-\ref{subsec:vem} have naturally explored adaptations of the computationally expensive E-step from the EM algorithm of Subsection~\ref{subsec:em}. As a result, they differ in their inference on the hidden variables of the generative model. Yet, the only hidden variable of interest is the conformation variable and associated molecular volumes. As a result, recent works have explored methods that avoid the inference on the ``nuisance" latent variables altogether. This is the approach taken by \acrfull{gan}.

\paragraph{Generative Adversarial Networks as Minimax Games} A generative adversarial model (GAN)\cite{Goodfellow2014GenerativeNets} is a method that estimates the parameters of a generative model, such as the one in Equation~\eqref{eq:image}, through an adversarial process. The GAN trains a ``generator" to produce images $x_i$ that best capture the training data distribution, while a discriminator estimates the probability that a given image $x_i$ came from the training data rather than the generator.

In other words, the generator and discriminator play the following two-player minimax game with value function $V$:
\begin{align*}
    V = \min_\theta \max_\phi 
    \mathbb{E}_{x \sim p_{\text{data}}(x)}\left[ 
            \log \text{Dis}_\phi(x)\right] 
    + \mathbb{E}_{h \sim p(h)} \left[
        \log \left(1 - \text{Dis}_\phi (f_{V_\theta}(h))\right) \right],
\end{align*}
where the generator is the cryo-EM generative model defined in Equation~\eqref{eq:image} and the discriminator has weights $\phi$ and is denoted by $\text{Dis}_\phi$. In this equation, $p_{\text{data}}(x)$ represents the probability distribution of the images, and $p(h)$ is a prior distribution on the hidden variables. The GAN training iterates two steps:
\begin{description}
 \item[(a) Discriminator - Update of parameters $\phi$], according to the gradient step:
    \begin{equation}
        \phi^{(t+1)} = \phi^{(t)} + \nabla_\phi
                \frac{1}{n}\sum_{i=1}^n \Big[ 
                \log \text{Dis}_\phi(x_i) + \log \left(1 - \text{Dis}_\phi (f_{V_\theta^{(t)}}(h_i))\right)
                \Big],
    \end{equation}  
 \item[(b) Generator - Estimation of model's parameter $\theta$], according to the gradient step:
     \begin{equation}
        \theta^{(t+1)} = \theta^{(t)} - \nabla_\theta
                \frac{1}{n}\sum_{i=1}^n \Big[ 
                \log \left(1 - \text{Dis}_{\phi^{(t)}} (f_{V_\theta}(h_i))\right)
                \Big],
    \end{equation}  
\end{description}
where in each Step (a) or (b), the "hidden variables" $h_i$ are sampled according to a prior distribution $p_h$. Even though a GAN iterates two steps, including one step related to the estimation of the model's parameters, its framework differ from the variations of the EM algorithm in the sense that the hidden variables are not inferred: it is enough to be able to randomly sample from them from some prior distribution.

\paragraph{GANs in \cryogan and \multicryogan} \cryogan uses a GAN to perform homogeneous reconstruction of molecular volumes, while using uniform sampling on the rotation, translation and CTF hidden variables. \multicryogan introduces heterogeneous reconstruction with this approach. Both \cryogan and \multicryogan use the Wasserstein variant \cite{wgan2017} of the traditional GANs.


\section*{Appendix D: Details on (Variational) Autoencoders}\label{appD-AEs}
\setcounter{subsection}{0}
\renewcommand{\thesubsection}{D.\arabic{subsection}}

Autoencoders (AEs) and variational autoencoders (VAEs) are the main realizations of the amortized variational inference approaches in the cryo-EM reconstruction literature. This appendix provides additional details to link the traditional presentations of the AEs and VAEs to the framework described in Section~\ref{sec2-inference}.
 
\subsection{Autoencoders}

An autoencoder traditionally aims to minimize the following loss function:
\begin{equation}
     \ell (\theta, \xi)
     = \sum_{i=1}^n  ||x_i -  \hat{x_i} ||^2 
     = \sum_{i=1}^n  ||x_i -  f_{V_\theta}(h_i)  ||^2 
     = \sum_{i=1}^n  ||x_i -  f_{V_\theta}(\text{Enc}_{\xi}(x_i) ||^2,
\end{equation}
where $f_{V_\theta}$ denotes here the output of the generative model (such as the one in Equation~\ref{eq:image}) and is usually called a reconstruction: $\hat x_i = f_{V_\theta}(h_i)$, while $\ell (\theta, \xi)$ is called the reconstruction loss. Note that the mean square error is used to quantify the reconstruction loss, but it can be replaced by other metrics. For example, the binary-cross entropy is a metric traditionally used to compare images $x_i$ and $\hat x_i$. In this context, the goal is to perform the double minization in $\theta$ and $\xi$ such that: 
\begin{equation}
  \hat \theta, \hat \xi = \argmin_{\theta, \xi} \ell(\theta, \xi). 
\end{equation}

By considering the decoder as latent variable model: $\hat x_i = f_{V_\theta}(h_i) + \epsilon_i$ with $\epsilon_i$ a standard Gaussian noise, this loss corresponds to the negative log-likelihood, informed by the encoder. 


To minimize this objective, the AE takes a gradient step at each iteration $t$, or backward pass through the network, such that:
\begin{description}
\item[(a) Encoder - Inference on latent variables $h_i$] The encoder updates its weight $\xi$ through a gradient step with learning rate $\alpha$:
\begin{equation}
    \xi^{(t)} = \xi^{(t-1)} - \alpha \nabla_\xi \ell(\theta, \xi).
\end{equation}
\item[(b) Decoder - Estimation of model's parameter $\theta$] The decoder updates its weight $\theta$ through a gradient step with learning rate $\alpha$:
\begin{equation}
    \theta^{(t)} = \theta^{(t-1)} - \alpha \nabla_\theta \ell(\theta, \xi).
\end{equation}
\end{description}
We note that this gradient descent is usually performed via stochastic gradient descent, such that only a mini-batch of the data is considered at each iteration to compute $\ell$, as opposed to the full dataset of $n$ images.


\subsection{Variational Autoencoders}

A variational autoencoder traditionally aims to minimize the following loss function, which is the negative ELBO \cite{Kingma2014Auto-EncodingBayes}:
\begin{equation}
    \mathcal{L}(\xi, \theta)  
    = \sum_{i=1}^n \mathbb{E}_{q_{\xi} (h_i|X_i))} (\log(p_\theta(X_i|z)) + KL ( q_{\xi}(h_i|X_i) || p(h_i)).
\end{equation}
In this loss, the first term is called the reconstruction term, estimated with one Monte Carlo sample through the so-called ``reparametrization trick", and is akin to the reconstruction loss of the AE. The second term is a KL divergence term that is called the regularization term, as it regularizes the posterior of the latent variable $h_i$ by forcing it to be close to the prior $p(h_i)$ of $h_i$ which is modeled by a standard Gaussian distribution.

We can rewrite this loss by using the functions $\text{Enc}_\xi$ and $\text{Dec}_\theta$, and assuming that the Monte Carlo sampling performed to compute the expectation happens by sampling a unique $\tilde h_i$ through $q_{\xi} (h_i|X_i)$ which is the convention adopted in these architectures:
\begin{equation}
    \mathcal{L}(\xi, \theta)  
    = \sum_{i=1}^n ||x_i - \hat{x_i} ||^2  - KL ( q_{\xi}(h_i|x_i) \parallel p(h_i)) \\
    = \sum_{i=1}^n ||x_i - \text{Dec}_{\theta}(\text{Enc}_{\xi}(x_i)||^2  - KL ( q_{\text{Enc}_\xi}(h_i|x_i) \parallel p(h_i)), \\
\end{equation}
where the KL term has a closed form in terms of the output of the encoder, due to the fact that the $q$ distribution belongs to a Gaussian family of diagonal covariance.

By realizing that the decoder only participates in the first term of the loss function, the VAE takes a gradient step at each iteration $t$, or backward pass through the network, such that:
\begin{description}
\item[(a) Encoder - Inference on latent variables $h_i$] The encoder updates its weight $\xi$ through a gradient step of learning rate $\alpha$ such that:
\begin{equation}
     \xi^{(t)} = \xi^{(t-1)} - \alpha \nabla_\xi \mathcal{L}(\xi, \theta) .
\end{equation}
\item[(b) Decoder - Estimation of model's parameter $\theta$] The decoder updates its weight $\theta$ through a gradient step with learning rate $\alpha$:
\begin{equation}
    \theta^{(t)} 
    = \theta^{(t-1)} - \alpha \nabla_\theta \mathcal{L}(\xi, \theta) 
    = \theta^{(t-1)} - \alpha \nabla_\theta \ell(\theta, \xi).
\end{equation}
 \end{description}
  

The main difference with the autoencoder is that the latent variable $h_i$ is considered as a random variable, rather than a fixed deterministic value. That is, $h_i$ is endowed with a parametric probability distribution represented by $q$ --- which is a Gaussian distribution with parameters output by the encoder, so that  $h_i \sim N(\mu_{\xi}(X_i),\sigma^2_{\xi}(X_i))$ and $\text{Enc}_\xi(X_i) = (\mu_xi(X_i), \sigma^2_\xi(X_i)$. Considering $h_i$s as random variables has been shown to lead superior reconstruction results over the autoencoder.


The VAE loss can be adapted as: $\mathcal{L}(\xi, \theta)  = \sum_{i=1}^n \mathbb{E}_{q_{\xi} (z|X_i))} (\log(p(X_i|z)) + \beta KL ( q_{\xi}(X_i|Z || p(x))$, where $\beta$ is an additional hyperparameter introduced in $\beta$-VAE\cite{Thiagarajan2017-VAE:Framework}. Traditional VAEs have $\beta=1$, to ensure that the actual negative ELBO is minimized. Yet, to prevent pathological issues in the fitting of VAEs (including posterior collapse), recent work has shown that counterbalancing the reconstruction error with the KL divergence through a $\beta$ could yield superior results.

In the cryo-EM implementations of the VAE architectures, the ELBO loss can be supplemented with additional terms that we name ``structure losses" for now. We will explain it in the next appendix as it depends on the structure of the decoder (i.e. exact choice of generative model) chosen by the method.

In practive, \cryoposenet, \cryoai and \egmm use an autoencoder, and \egmm additionally implements an variational autoencoder with a variational family of Gaussian distributions with fixed isotropic variance $\sigma_0$. The AE architecture is used in \cryoposenet with a traditional L2 reconstruction loss, in \cryoai with a ``symmetrized" L2 reconstruction loss and in \egmm with a tailored reconstruction loss that relies on the Fourier ring correlation (FRC) reconstruction metric.

\cryovaegan, \cryodrgn, \cryofold, and \cryodeepmind use VAEs with a variational family of Gaussian distributions with diagonal covariance matrix, to respectively infer $(R_i, \text{CTF}_i)$ (CryoVAEGAN), $z_i$ (CryoDRGN and CryoFold) and $(z_i, R_i)$ (atomVAE) -- see Table~\ref{tab:inference}.


The VAE architecture is used in \cryodrgn with the negative ELBO loss, and in \cryovaegan with the negative ELBO loss extended with the $\beta$ hyper-parameter described in the supplementary materials, an additional geometric regularization term, while the reconstruction loss relies on the binary cross-entropy as opposed to the L2 reconstruction loss. \cryodeepmind also implements a VAE with a modified ELBO loss that relies on important sampling (not detailed in this review), that leverages the $\beta$ hyperparameter and additionally includes a ``structure loss" --- see supplementary materials.

\section*{Appendix E: Constraints}\label{appE-constraints}
\setcounter{subsection}{0}
\renewcommand{\thesubsection}{E.\arabic{subsection}}

As described in the introduction and following the exposition by Scheres \cite{Scheres2012RELION} (2012), the molecule reconstruction problem is a difficult, highly non-linear inverse problem, which makes the parametrization of the shape and associated constraints particularly important. This parametrization amounts to impose structure on the desired reconstructed shapes --- either by leveraging domain knowledge on the properties of molecular volumes, or by using external information to guide the reconstruction.

\textit{From a physics/biology perspective}, this can be seen as a necessary enrichment of the cryo-EM data with either external assumptions and/or domain knowledge on the properties of the solution to ensure a more accurate recovery. \textit{From a statistics perspective}, formulated this way, the problem rapidly takes on a Bayesian flavour, and the objective of this step is to find the right ``prior" on the distribution of our latent variables. \textit{From a computational perspective}, this corresponds to adding additional terms to the loss/objective defining the optimization problem, and effectively explains what we term ``structure loss" in Section~\ref{sec2-inference}. We detail here the additional constraints that can equip the inference methods across the reconstruction algorithms.


\subsection{Smoothness}

Smoothness refers to the property by which a signal, or quantity of interest, varies with "no abrupt change" over continuous regions. As explained by Scheres (2012) \cite{Scheres2012RELION}, "because macromolecules consist of atoms that are connected through chemical bonds, the scattering potential will vary smoothly in space, especially at less than atomic resolution."  Smoothness of the recovered scattering potential $V$ thus appears to be a reasonable assumption, which is implemented in different ways depending on the parametrization chosen:

\begin{description}
\item[Smoothness of the 3D image.] The smoothness assumption holds in image space as the image is a projection of the electron field --- which is itself continuous. This smoothness is encoded by the normal distribution. This also calls to mind the "ridge penalty", a similar type of regularisation in statistics.

\item[Smoothness of the Fourier coefficients.]   The smoothness of the 3D density map translates into smoothness over neighbouring Fourier coefficients. In a Bayesian pipeline, this is typically parametrized by assuming that these coefficients are sampled from a normal distribution.  
    The RELION algorithm (Scheres \cite{Scheres2012RELION}) is based on such a smoothness assumption.  This prior is encoded by assuming independent Gaussian priors on the Fourier components of the signal: $ V_l \sim N(0, \tau_l^2) $. Note here that the algorithm is not encouraging any other type of structure (e.g sparsity through spike-and-slab prior, etc).
    
\item[Smoothness of the deformation field.] 3DFlex \cite{Punjani2021} directly exploits the knowledge that conformational variability of a protein is the result of physical processes that transport density over space.  This means that mass and local geometry are preserved. As a result, this method implements a convection operator that outputs the deformation field. 3DFlex exploits prior knowledge of smoothness and local rigidity in the deformation field. 

\item[Smoothness of the function over 3D coordinates in Fourier domain.] Cryo-DRGN\cite{Zhong2019ReconstructingModels} and Cryo-Fold\cite{zhong2021exploring} represent the volume as the function $f: \Omega^3 \rightarrow \mathbb{R}$ over a 3D domain. In practice, this function is implemented by a neural network, which constrains it to be continuous, and possibly smooth if the activation functions used by the network are themselves smooth (sigmoids, for example).
\end{description}

\subsection{Rigidity}

Other constraints exploit physical properties and knowledge of the system to constrain the reconstruction. Specifically, due to the fact that the molecules studied in cryo-EM are frozen, we can assume that their conformational heterogeneity will not present large variations. As a result, we can assume that the molecular structure has some type of ``rigidity". This is implemented in practice by refraining the variables describing the shape heterogeneity from varying excessively. Depending on the parametrization chosen to represent the volume heterogeneity, the rigidity constraint takes different forms:

\begin{description}
\item[Rigidity of the deformation field.] The deformation field that parametrizes volume heterogeneity in 3D-Flex\cite{Punjani2021} can be constrained to only generate ``small" deformation, through a regularization term. This is integrated as a ``structure loss" of the deep learning training procedure of 3D-Flex\cite{Punjani2021}.
    
\item[Rigidity in (pseudo)-atoms coordinates.] The (pseudo-) atoms coordinates that parametrize the volume heterogeneity in Cryo-Fold\cite{zhong2021exploring} are constrained to be close to a base conformation's coordinates, via an $L2$ penalization on the deformations. This forms the ``structure loss" integrated in the objective function for the training procedure.

\item[Rigidity in deviation from (pseudo)-atoms coordinates.] Here, the volume is parametrized as a set of deviations of pseudo-atoms coordinates, compared to a base conformation, as in Cryo-DeepMind\cite{Rosenbaum2021}. The rigidity of the molecular structure is thus enforced by adding a constraint that takes the form of a L2 regularization constraining the $\Delta c_j$ to be small, and effectively forming the ``structure loss" added to the optimization objective of this approach.
\end{description}

We also note that the methods that do not assume any heterogeneity in volumes $V_i$ but rather model the volume as a unique possible conformation $V$, are essentially relying on the rigidity assumption, assuming that variations around $V$ are not large.

\section*{Appendix F: Experiments}\label{appF-experiments}
This appendix summarizes the experiments conducted in the papers cited in this review, in Table~\ref{tab:experiments}. As mentioned in the main text, we observe a great diversity of datasets, which explains the difficulty encountered in comparing methods' performances and accuracies.

\begin{table}[]
\scalebox{0.91}{
\begin{tabular}{|l|l|l|l|l|l|}
\hline
\rowcolor[HTML]{EFEFEF} 
                                                           & Biomolecule                             & \# of images & Image size & Data Type    & Noise  \\ \hline
\rowcolor[HTML]{FAFAE7} 
\cellcolor[HTML]{EFEFEF}                                   & \cellcolor[HTML]{FAFAE7}                & 50,000       & 128 x 128  & Synthetic    & No/Yes \\ \cline{3-6} 
\rowcolor[HTML]{FAFAE7} 
\cellcolor[HTML]{EFEFEF} &
  \multirow{-2}{*}{\cellcolor[HTML]{FAFAE7}Ribosome 80S (EMPIAR-10028)} &
  105,247 &
  90 x 90 &
  Experimental &
  Yes \\ \cline{2-6} 
\rowcolor[HTML]{FAFAE7} 
\cellcolor[HTML]{EFEFEF}                                   & Ribosome 50S (EMPIAR-10076)             & 131,899      & 90 x 90    & Experimental & Yes    \\ \cline{2-6} 
\rowcolor[HTML]{FAFAE7} 
\multirow{-4}{*}{\cellcolor[HTML]{EFEFEF}\cryodrgn}        & Protein complex                         & 50,000       & 64 x 64    & Synthetic    & N/A    \\ \hline
\rowcolor[HTML]{FAFAE7} 
\cellcolor[HTML]{EFEFEF}\cryofold                          & Haemoglobin (PDB 5NI1)                  & 50,000       & 128 x 128  & Synthetic    & Yes    \\ \hline
\rowcolor[HTML]{FAFAE7} 
\cellcolor[HTML]{EFEFEF}                                   & Tri-snRNP spliceosome (EMPIAR-10073)    & 102,500      & 180 x 180  & Experimental & Yes    \\ \cline{2-6} 
\rowcolor[HTML]{FAFAE7} 
\multirow{-2}{*}{\cellcolor[HTML]{EFEFEF}\threedflex} &
  TRPV1 ion-channel (EMPIAR-10059) &
  200,000 &
  128 x 128 &
  Experimental &
  Yes \\ \hline
\rowcolor[HTML]{FEF5F4} 
\cellcolor[HTML]{EFEFEF}\cryoposenet                       & E.coli adenylate kinase (PDB 4AKE1)     & 9,000        & 128 x 128  & Synthetic    & No/Yes \\ \hline
\rowcolor[HTML]{FAFAE7} 
\cellcolor[HTML]{EFEFEF}                                   & Ribosome 50S (EMPIAR-10076)             & 124,900      & N/A        & Experimental & Yes    \\ \cline{2-6} 
\rowcolor[HTML]{FAFAE7} 
\cellcolor[HTML]{EFEFEF}                                   & Precatalytic spliceosome (EMPIAR-10180) & 327,490      & N/A        & Experimental & Yes    \\ \cline{2-6} 
\rowcolor[HTML]{FAFAE7} 
\multirow{-3}{*}{\cellcolor[HTML]{EFEFEF}\egmm}            & SARS-CoV-2 spike protein (EMPIAR-10492) & 55,159       & N/A        & Experimental & Yes    \\ \hline
\rowcolor[HTML]{FEF5F4} 
\cellcolor[HTML]{EFEFEF}\cryomaxwelling                    & GroEL-GroES protein                     & 40,000       & 128 x 128  & Synthetic    & Yes    \\ \hline
\rowcolor[HTML]{FEF5F4} 
\cellcolor[HTML]{EFEFEF}                                   & \cellcolor[HTML]{FEF5F4}                & 2, 544       & 128 x 128  & Synthetic    & No     \\ \cline{3-6} 
\rowcolor[HTML]{FEF5F4} 
\multirow{-2}{*}{\cellcolor[HTML]{EFEFEF}\cryovaegan} &
  \multirow{-2}{*}{\cellcolor[HTML]{FEF5F4}Ribosome 80S (EMPIAR-10028)} &
  5,119 - 8,278 - 4,917 &
  128 x 128 &
  Experimental &
  Yes \\ \hline
\rowcolor[HTML]{FAFAE7} 
\cellcolor[HTML]{EFEFEF}\cryodeepmind                      & Aurora A Kinase (Simulated)             & 63,000       & 64 x 64    & Synthetic    & Yes    \\ \hline
\rowcolor[HTML]{FAFAE7} 
\cellcolor[HTML]{FAFAE7}                                   & 80S ribosome (PDB 3J79 and 3J7A)        & 1,000,000    & 128 x 128  & Synthetic    & No/Yes \\ \cline{2-6} 
\rowcolor[HTML]{FAFAE7} 
\cellcolor[HTML]{FAFAE7}                                   & SARS-CoV-2 spike protein (PDB 6VYB)     & 1,000,000    & 128 x 128  & Synthetic    & No/Yes \\ \cline{2-6} 
\rowcolor[HTML]{FAFAE7} 
\cellcolor[HTML]{FAFAE7}                                   & Spliceosome (PDB 5NRL)                  & 1,000,000    & 128 x 128  & Synthetic    & No/Yes \\ \cline{2-6} 
\rowcolor[HTML]{FAFAE7} 
\multirow{-4}{*}{\cellcolor[HTML]{FAFAE7}\cryoai}          & 80S ribosome (EMPIAR-10028)             & 105,247      & 256 x 256  & Experimental & Yes    \\ \hline
\rowcolor[HTML]{FEF5F4} 
\cellcolor[HTML]{EFEFEF}\cryogan                           & $\beta$-galactosidase                   & 41,000       & 180 x 180  & Synthetic    & Yes    \\ \hline
\rowcolor[HTML]{FEF5F4} 
\cellcolor[HTML]{EFEFEF}                                   & $\beta$-galactosidase (EMPIAR-10061)    & 41,123       & 192 x 192  & Experimental & Yes    \\ \hline
\rowcolor[HTML]{FAFAE7} 
\cellcolor[HTML]{EFEFEF}\multicryogan & Heat-shock protein Hsp90                & 100,000      & 32 x 32    & Synthetic    & Yes    \\ \hline
\end{tabular}
}
\caption{Summary of synthetic and experimental cryo-EM datasets on which the reconstruction methods have been tested. The light red background indicates methods that perform homogeneous reconstruction, and the light yellow background heterogeneous reconstructions.}
\label{tab:experiments}
\end{table}

\section*{Appendix G: Classification of Reconstruction Methods}\label{appG-classification}

This last appendix provides details in the classification of the reconstruction methods, in the form of two tables. Table~\ref{tab:volumes} compares the generative models and Table~\ref{tab:inference} compares the inference methods. 

\begin{table}[h!]
\centering
\resizebox{\textwidth}{!}{
\begin{tabular}{|c|c|
>{\columncolor[HTML]{D9EAFF}}c |
>{\columncolor[HTML]{D9EAFF}}c |
>{\columncolor[HTML]{D9EAFF}}c |
>{\columncolor[HTML]{D9EAFF}}c |
>{\columncolor[HTML]{D9EAFF}}c |
>{\columncolor[HTML]{D9EAFF}}c |}
\hline
    \multicolumn{2}{|c|}{\cellcolor[HTML]{EFEFEF}
        \textbf{\begin{tabular}[c]{@{}c@{}}Volume\\Param.\end{tabular}}} & 
    \cellcolor[HTML]{EFEFEF}
        \textbf{Space} &
    \cellcolor[HTML]{EFEFEF}\textbf{Conformational Model} &
    \cellcolor[HTML]{EFEFEF}
        \textbf{Approach} &
    \cellcolor[HTML]{EFEFEF}
        \textbf{\begin{tabular}[c]{@{}c@{}}Hidden\\ Variable\end{tabular}} &
    \cellcolor[HTML]{EFEFEF}
        \textbf{\begin{tabular}[c]{@{}c@{}}Known\\ Variable\end{tabular}} &
    \cellcolor[HTML]{EFEFEF}
        \textbf{\begin{tabular}[c]{@{}c@{}}Reference\\ Volume\end{tabular}}\\ 
\hline
    \hline
        & 
        & 
        \cellcolor[HTML]{D9EAFF} & 
        \cellcolor[HTML]{D9EAFF} & 
        \cryoposenet & 
        \begin{tabular}[c]{@{}c@{}}-\\Rotation\\-\end{tabular} & 
        \begin{tabular}[c]{@{}c@{}}CTF\\-\\Translation\end{tabular} & 
        Free\\ 
    \cline{5-8} 
        & 
        &
        \multirow{-2}{*}{\cellcolor[HTML]{D9EAFF}Image} & 
        \multirow{-2}{*}{\cellcolor[HTML]{D9EAFF}
            \begin{tabular}[c]{@{}c@{}}Homogeneous\\ $z \longrightarrow V$ \end{tabular}} &
        \cryogan &
        \begin{tabular}[c]{@{}c@{}}-\\Rotation\\Translation\end{tabular} & 
        \begin{tabular}[c]{@{}c@{}}CTF\\-\\-\end{tabular} & 
        Free\\
    \cline{4-8} 
        \parbox[c]{2mm}{\multirow{-3}{*}{\rotatebox[origin=c]{90}{\textbf{Discrete Representation}}}} & 
        \parbox[c]{2mm}{\multirow{-2}{*}{\rotatebox[origin=c]{90}{Voxel Grid}}} & 
        & 
        \begin{tabular}[c]{@{}c@{}}Heterogeneous\\ $z \longrightarrow V$ \end{tabular} & 
        \multicryogan & 
        \begin{tabular}[c]{@{}c@{}}-\\Rotation\\Translation\end{tabular}& 
        \begin{tabular}[c]{@{}c@{}}CTF\\-\\-\end{tabular} &
        Free\\ 
  \cline{4-8}
        & & &
        \begin{tabular}[c]{@{}c@{}}Heterogeneous\\ $z \longrightarrow f(U(z_j), V_0)$ \end{tabular} & 
        \threedflex & 
        \begin{tabular}[c]{@{}c@{}}-\\-\\-\end{tabular}& 
        \begin{tabular}[c]{@{}c@{}}CTF\\Rotation\\Translation\end{tabular} &
        Yes\\ 
    \cline{3-8} 
        &
        & 
        \cellcolor[HTML]{DFFFD3} & 
        \cellcolor[HTML]{DFFFD3}
            \begin{tabular}[c]{@{}c@{}}Homogeneous\\ $z = (\mu_V, \sigma_V) \longrightarrow V$\end{tabular}& 
        \cellcolor[HTML]{DFFFD3}\cryomaxwelling & 
        \cellcolor[HTML]{DFFFD3}
            \begin{tabular}[c]{@{}c@{}}-\\Rotation\\Translation\end{tabular}&
        \cellcolor[HTML]{DFFFD3}
            \begin{tabular}[c]{@{}c@{}}CTF\\-\\-\end{tabular} &
        \cellcolor[HTML]{DFFFD3}Free\\
    \cline{4-8} 
        & 
        & 
        \multirow{-2}{*}{\cellcolor[HTML]{DFFFD3}Fourier} & 
        \cellcolor[HTML]{DFFFD3}
            \begin{tabular}[c]{@{}c@{}}Heterogeneous\\ $z \in \{1, .., K\} \longrightarrow \{V_1, ..., V_K\}$ \end{tabular}& 
        \cellcolor[HTML]{DFFFD3}\relion &
        \cellcolor[HTML]{DFFFD3}
            \begin{tabular}[c]{@{}c@{}}-\\Rotation\\Translation\end{tabular}&
        \cellcolor[HTML]{DFFFD3}
            \begin{tabular}[c]{@{}c@{}}CTF\\-\\-\end{tabular} &
        \cellcolor[HTML]{DFFFD3}Free\\ 
    \cline{4-8} 
        & 
        & 
        \multirow{-2}{*}{\cellcolor[HTML]{DFFFD3}} & 
        \cellcolor[HTML]{DFFFD3}
            \begin{tabular}[c]{@{}c@{}}Heterogeneous\\ $z \in \{1, .., K\} \longrightarrow \{V_1, ..., V_K\}$ \end{tabular}& 
        \cellcolor[HTML]{DFFFD3}\cryosparc &
        \cellcolor[HTML]{DFFFD3}
            \begin{tabular}[c]{@{}c@{}}-\\Rotation\\Translation\end{tabular}&
        \cellcolor[HTML]{DFFFD3}
            \begin{tabular}[c]{@{}c@{}}CTF\\-\\-\end{tabular} &
        \cellcolor[HTML]{DFFFD3}Free\\ 
    \hline
    \hline
        & 
        \parbox[t]{2mm}{}{\multirow{-1}{*}{\rotatebox[origin=c]{90}{\begin{tabular}[c]{@{}c@{}}Neural\\ Network\end{tabular}}}} & 
        Image & 
        \begin{tabular}[c]{@{}c@{}}Homogeneous\\ $z = (\mu_V, \sigma_V, CTF_V) \longrightarrow V_i(2D)$\end{tabular} & 
        \cryovaegan & 
        \begin{tabular}[c]{@{}c@{}}CTF\\2D Rotation\\-\end{tabular} & 
        \begin{tabular}[c]{@{}c@{}}-\\-\\Translation\end{tabular} &
        Free\\ 
    \cline{3-8} 
        \parbox[t]{2mm}{\multirow{-2}{*}{\rotatebox[origin=c]{90}{\textbf{Continuous Field}}}} & 
        & 
        \cellcolor[HTML]{DFFFD3} & 
        \cellcolor[HTML]{DFFFD3}
            \begin{tabular}[c]{@{}c@{}}Heterogeneous\\ $z \longrightarrow V_j$\end{tabular} & 
        \cellcolor[HTML]{DFFFD3}\cryodrgn & 
        \cellcolor[HTML]{DFFFD3}
            \begin{tabular}[c]{@{}c@{}}-\\Rotation\\Translation\end{tabular}&
        \cellcolor[HTML]{DFFFD3}
            \begin{tabular}[c]{@{}c@{}}CTF\\-\\-\end{tabular} &   
        \cellcolor[HTML]{DFFFD3}Free\\ 
        \cline{4-8}
        & &
        \cellcolor[HTML]{DFFFD3}Fourier & 
        \cellcolor[HTML]{DFFFD3}
            \begin{tabular}[c]{@{}c@{}}Homogeneous\\ $z \longrightarrow V_j$\end{tabular} & 
        \cellcolor[HTML]{DFFFD3}\cryoai & 
        \cellcolor[HTML]{DFFFD3}
            \begin{tabular}[c]{@{}c@{}}-\\Rotation\\Translation\end{tabular}&
        \cellcolor[HTML]{DFFFD3}
            \begin{tabular}[c]{@{}c@{}}CTF\\-\\-\end{tabular} &   
        \cellcolor[HTML]{DFFFD3}Free\\ 
    \cline{2-8}
        &
        &
        &
        \cellcolor[HTML]{D9EAFF}
            \begin{tabular}[c]{@{}c@{}}Heterogeneous\\ $z \longrightarrow \{c_j, A_j, \sigma_j\}_{j\in[1,N]}$ or $\{\Delta c_j\}_{j\in[1,N]}$ ; ($N$ ``blobs") \end{tabular}& 
        \egmm & 
        \cellcolor[HTML]{D9EAFF}
            \begin{tabular}[c]{@{}c@{}}-\\-\\-\end{tabular}& 
        \cellcolor[HTML]{D9EAFF}
            \begin{tabular}[c]{@{}c@{}}CTF\\Rotation\\Translation\end{tabular} &
        \cellcolor[HTML]{D9EAFF}Free or Yes\\ 
    \cline{4-8}
        &
        \parbox[t]{2mm}{}{\multirow{-2}{*}{\rotatebox[origin=c]{90}{\begin{tabular}[c]{@{}c@{}}Gaussian\\ Mixture\end{tabular}}}} &
        \multirow{-2}{*}{\cellcolor[HTML]{D9EAFF}Image} &
        \cellcolor[HTML]{D9EAFF}
            \begin{tabular}[c]{@{}c@{}}Heterogeneous\\ $z \longrightarrow \{c_j\}_{j\in[1,N]}$; ($N$ residues) \end{tabular}& 
        \cryofold& 
        \cellcolor[HTML]{D9EAFF}
            \begin{tabular}[c]{@{}c@{}}-\\-\\-\end{tabular}& 
        \cellcolor[HTML]{D9EAFF}
            \begin{tabular}[c]{@{}c@{}}CTF\\Rotation\\Translation\end{tabular} &
        \cellcolor[HTML]{D9EAFF}Free\\ 
    \cline{4-8}
        &
        &
        &
        \cellcolor[HTML]{D9EAFF}
            \begin{tabular}[c]{@{}c@{}}Heterogeneous\\ $z \longrightarrow \{\Delta c_j\}_{j\in[1,N]}$; ($N$ residues) \end{tabular}& 
        \cryodeepmind& 
        \cellcolor[HTML]{D9EAFF}
            \begin{tabular}[c]{@{}c@{}}-\\Rotation\\Translation\end{tabular}& 
        \cellcolor[HTML]{D9EAFF}
            \begin{tabular}[c]{@{}c@{}}CTF\\-\\-\end{tabular} &
        \cellcolor[HTML]{D9EAFF}Yes\\ 
    \hline
\end{tabular}
}
\caption{Classification of reconstruction methods in terms of the parametrization of the volume $V$ (or $\tilde{V}$), \textit{i.e.} the model of conformational heterogeneity with the conformation variable $z$, use of image versus Fourier space, representation as a discrete or continuous field, and with a reference-free or reference-based encoding. We also indicate whether nuisance variables are hidden or assumed to be known in the reconstruction methods cited. When not specified, the latent variable $z$ belongs to a vector space $\mathbb{R}^L$.} 
\label{tab:volumes}
\end{table}

\begin{table}[]
\scalebox{0.98}{
\begin{tabular}{|c|c|cc|cc|}
\hline
\rowcolor[HTML]{DAE8FC} 
\cellcolor[HTML]{EFEFEF} &
  \textbf{Posterior $p(h_i|x_i)$} &
  \multicolumn{2}{c|}{\cellcolor[HTML]{DAE8FC}\textbf{Mode $\hat h_i$}} &
  \multicolumn{2}{c|}{\cellcolor[HTML]{DAE8FC}\textbf{Variational $q(h_i)$}} \\ \hline
\rowcolor[HTML]{EFEFEF} 
 &
  \textbf{} &
  \multicolumn{1}{c|}{\cellcolor[HTML]{EFEFEF}\textbf{Non-Amortized}} &
  \textbf{\begin{tabular}[c]{@{}c@{}}Amortized \\ (encoder)\end{tabular}} &
  \multicolumn{1}{c|}{\cellcolor[HTML]{EFEFEF}\textbf{Non-Amortized}} &
  \textbf{\begin{tabular}[c]{@{}c@{}}Amortized\\ (encoder)\end{tabular}} \\ \hline
Distribution on $h_i$ &
  $p(h_i|x_i)$ &
  \multicolumn{1}{c|}{$ \argmax _z p(h_i|x_i)$} &
  $\text{Enc}_\xi(x_i)$ &
  \multicolumn{1}{c|}{$q_{\eta_i}  \in \mathcal{Q}$} &
  $q_{\text{Enc}_\xi(x_i)}\in \mathcal{Q}$ \\ \hline
\rowcolor[HTML]{EFEFEF} 
\multicolumn{1}{|l|}{\cellcolor[HTML]{EFEFEF}\textbf{Name}} &
  \textbf{EM} &
  \multicolumn{1}{c|}{\cellcolor[HTML]{EFEFEF}\textbf{Approx. EM}} &
  \textbf{Am. EM / AE} &
  \multicolumn{1}{c|}{\cellcolor[HTML]{EFEFEF}\textbf{Var. EM  / AD}} &
  \textbf{Var. Am. EM / VAE} \\ \hline
\cellcolor[HTML]{EFEFEF} &
  \begin{tabular}[c]{@{}c@{}}RELION\cite{Scheres2012RELION}\\ (pose $R_i, t_i$ and \\ conformation $z_i$)\end{tabular} &
  \multicolumn{1}{c|}{\begin{tabular}[c]{@{}c@{}}RELION\cite{Scheres2012RELION}\\ (noise $\sigma_i$) (*)\end{tabular}} &
  \begin{tabular}[c]{@{}c@{}}CryoPoseNet \cite{Nashed2021endtoend} \\ (rotation $R_i$)\end{tabular} &
  \multicolumn{1}{c|}{\begin{tabular}[c]{@{}c@{}}3DFlex\cite{Punjani2021}\\ (conformation $z_i$)\end{tabular}} &
  \begin{tabular}[c]{@{}c@{}}CryoDRGN\cite{Zhong2019ReconstructingModels}\\ (conformation $z_i$)\end{tabular} \\ \cline{2-6} 
\cellcolor[HTML]{EFEFEF} &
   &
  \multicolumn{1}{c|}{\begin{tabular}[c]{@{}c@{}}\cryosparc\\ (rotation $R_i$) (*)\end{tabular}} &
  \begin{tabular}[c]{@{}c@{}}E2GMM\cite{Chen2021}\\ (conformation $z_i$)\end{tabular} &
  \multicolumn{1}{c|}{\begin{tabular}[c]{@{}c@{}}\cryomaxwelling\\ (pose $R_i, t_i$ and \\ conformation $z$)\end{tabular}} &
  \begin{tabular}[c]{@{}c@{}}CryoVAEGAN\cite{MiolanePoitevin2020CVPR}\\ (2D rotation $R_i$\\ and $\text{CTF}_i$)\end{tabular} \\ \cline{2-6} 
\cellcolor[HTML]{EFEFEF} &
   &
  \multicolumn{1}{c|}{\begin{tabular}[c]{@{}c@{}}CryoDRGN\cite{Zhong2019ReconstructingModels}\\ (pose $R_i, t_i$) (*)\end{tabular}} &
  \begin{tabular}[c]{@{}c@{}}\cryoai \\ (pose $R_i, t_i$)\end{tabular} &
  \multicolumn{1}{c|}{} &
  \begin{tabular}[c]{@{}c@{}}atomVAE\cite{Rosenbaum2021}\\ (pose $R_i, t_i$ and \\ conformation $z_i$)\end{tabular} \\ \cline{2-6} 
\multirow{-4}{*}{\cellcolor[HTML]{EFEFEF}\begin{tabular}[c]{@{}c@{}}Inference on $h_i$\\ (E-step)\end{tabular}} &
  \multicolumn{1}{l|}{} &
  \multicolumn{1}{c|}{\begin{tabular}[c]{@{}c@{}}3DFlex\cite{Punjani2021}\\ (conformation $z_i$)\end{tabular}} &
  \multicolumn{1}{l|}{} &
  \multicolumn{1}{l|}{} &
  \begin{tabular}[c]{@{}c@{}}CryoFold\cite{zhong2021exploring}\\ (conformation $z_i$)\end{tabular} \\ \hline
\end{tabular}
}
\caption{Classification of cryo-EM reconstruction algorithms with respect to the inference on the conformation $z_i$ and on the unknown hidden variables $h_i$. The inference step updates the posterior $p(h_i|x_i$, or its mode $\hat h_i$, or a variational approximation $q(h_i)$ of it --- which corresponds to the 3 main columns of the table. The works by Zhong et al. (2019, 2021) are classified in several cells, as inference of the conformation $z_i$ is performed via a VAE, but the inference of the poses $R_i, t_i$ is performed with an optimization algorithm. The notation (*) specifies that the optimization associated with this step is fully performed (until convergence to local extremum) while its absence indicates that only a gradient step is taken toward the optimum.}
\label{tab:inference}
\end{table}

\end{document}